\documentclass[onefignum,onetabnum,final]{siamart171218}

\usepackage{tikz}
\usepackage{tikz-3dplot}
\usepackage[T1]{fontenc}
\usepackage{lineno}
\usepackage{epstopdf}
\usepackage{pifont}
\usepackage{dsfont}
\usepackage{latexsym}
\usepackage{graphics}
\usepackage{graphicx}
\usepackage{float}
\usepackage[all]{xy}
\usepackage{ifpdf}
\usepackage{multirow}
\usepackage{amssymb, amsmath}
\usepackage{hhline}
\usepackage{centernot}
\usepackage{verbatim}
\usepackage{enumitem}
\usepackage{setspace}
\usepackage{lipsum}
\usepackage[caption=false]{subfig}

\usepackage{mathtools}
\usepackage[normalem]{ulem}
\usepackage{xcolor} 
\colorlet{BLUE}{blue}
\colorlet{RED}{red}
\colorlet{GREEN}{green}
\colorlet{GRAY}{gray}

\usepackage[nice]{nicefrac}

\newsiamremark{remark}{Remark}
\newsiamremark{example}{Example}
\newsiamthm{fact}{Fact}
\newsiamthm{conjecture}{Conjecture}

\newcommand{\ba}{\begin{align}}
\newcommand{\ea}{\end{align}} 
\newcommand{\be}{\begin{equation}}
\newcommand{\ee}{\end{equation}}

\newcommand{\bea}{\begin{eqnarray}}
\newcommand{\eea}{\end{eqnarray}}
\newcommand{\barr}{\begin{array}}
\newcommand{\earr}{\end{array}}
\newcommand{\bn}{\begin{enumerate}}
\newcommand{\en}{\end{enumerate}}
\newcommand{\bi}{\begin{itemize}}
\newcommand{\ei}{\end{itemize}}
\newcommand{\bbbm}{\begin{pmatrix}}
\newcommand{\eeem}{\end{pmatrix}}

\newcommand{\cH}{{\cal H}}

\newcommand{\cP}{{\cal P}}

\newcommand{\E}{{\mathbb E}}

\newcommand{\R}{{\mathbb R}}

\newcommand{\bt}{\beta}

\newcommand{\ga}{\gamma}

\newcommand{\ep}{\epsilon}
\newcommand{\ka}{\kappa}

\newcommand{\ignore}[1]{}{}
\newcommand{\noin}{\noindent}

\newcommand{\nn}{\nonumber}

\newcommand{\q}{\quad}

\newcommand{{\QED}}{{\hfill QED} \bigskip}

\renewcommand{\subset}{\subseteq}

\DeclareMathOperator{\conv}{conv}

\definecolor{darkspringgreen}{rgb}{0.09, 0.45, 0.27} 
\definecolor{darkgray}{rgb}{0.66, 0.66, 0.66}

\numberwithin{equation}{section}
\numberwithin{theorem}{section}

\tikzset{
  partial ellipse/.style args={#1:#2:#3}{
      insert path={+ (#1:#3) arc (#1:#2:#3)}
 }
}

\renewcommand\theenumi{(\roman{enumi})}
\renewcommand\theenumii{(\alph{enumii})}

\headers{Dimension Reduction in VMOT}{J. Hiew, T. Lim, B. Pass, and M. Souza}

\title{Dimension Reduction in Martingale Optimal Transport: Geometry and Robust Option Pricing \thanks{All authors have contributed equally and are listed alphabetically. T.L. wishes to express gratitude to the Korea Institute of Advanced Study (KIAS) AI research group and the director Hyeon, Changbong for their hospitality and support during his stay at KIAS in 2023, where parts of this work were performed.  B.P. is pleased to acknowledge the support of Natural Sciences and Engineering Research Council of Canada Discovery Grant number  04658-2018. The work of J.H. and M.S. is in partial fulfillment of their doctoral degrees.}}

\author{Joshua Zoen-Git Hiew\thanks{Department of Mathematical and Statistical Sciences, University of Alberta, Edmonton AB Canada (\email{joshuazo@ualberta.ca}).}
\and Tongseok Lim\thanks{Mitchell E. Daniels, Jr. School of Business, Purdue University, West Lafayette, Indiana 47907, USA (\email{lim336@purdue.edu}).}
\and Brendan Pass\thanks{Department of Mathematical and Statistical Sciences, University of Alberta, Edmonton AB Canada (\email{pass@ualberta.ca}).}
\and Marcelo Cruz de Souza\thanks{Financial System Monitoring Department, Central Bank of Brazil, Fortaleza, CE, Brazil (\email{marcelo.souza@bcb.gov.br}).}
}

\begin{document}

\maketitle

\begin{abstract} 
This paper addresses the problem of robust option pricing within the framework of Vectorial Martingale Optimal Transport (VMOT). We investigate the geometry of VMOT solutions for $N$-period market models and demonstrate that, when the number of underlying assets is $d=2$ and the payoff is sub- or supermodular, the extremal model reduces to a single-factor structure in the first period. This structural result allows for a significant dimension reduction, transforming the problem into a more tractable format. We prove that this reduction is specific to the two-asset case and provide counterexamples showing it generally fails for $d \geq 3$. Finally, we exploit this monotonicity to develop a reduced-dimension Sinkhorn algorithm. Numerical experiments demonstrate that this structure-preserving approach reduces computational time by approximately 99\% compared to standard methods while improving accuracy.
\end{abstract}



\section{Introduction}\label{sec:Introduction}

  We denote $[n]:= \{1, 2, ..., n\}$ for $n \in \mathbb{N}$, and let $\cP(\Omega)$ denote the set of all probability measures (distributions) over a set $\Omega$. Fix $d, N \in \mathbb{N}$. Let $\vec \mu = (\vec\mu_1, \dots, \vec\mu_N)$ such that $\vec\mu_t = (\mu_{t,1}, \dots, \mu_{t,d})$ for $t \in [N]$, denoting vectors of probability measures (called marginals) on $\mathbb{R}$. Throughout the paper, we assume that all distributions have finite first moments.
  
  We consider the following space of \emph{Vectorial Martingale Transports} (VMT) from $\vec \mu_t$ to $\vec \mu_{t+1}$, $t = 1, \dots, N-1$ (see \cite{Lim22, Lim23}):
  \begin{align*}
    {\rm VMT}(\vec\mu) := \{ \pi \in \ &\cP(\R^{Nd}) \ | \  \pi = {\rm Law} (X), \, X = (X_1,\dots,X_N), \\
    &\E_\pi[X_{t+1}|X_1, \dots, X_t]=X_t, \, {\rm Law}(X_{t,i}) =\mu_{t,i}, \, \text{ for all } t \in [N], \, i \in [d] \}, \nn
  \end{align*}
  where $X = (X_1, \dots, X_N) \in \mathbb{R}^{Nd}$ are random vectors. 
  
  For a distribution $\pi \in \cP(\mathbb{R}^{Nd})$, we denote by $\pi^t \in \cP(\mathbb{R}^d)$ its $t^{\rm th}$ period marginal, i.e., if $\pi = {\rm Law} (X)$, then $\pi^t = {\rm Law} (X_t)$.
  We also denote by $\Pi(\vec\mu_t)$ the set of couplings of the marginals $\mu_{t,i}$, i.e.,
  \begin{align*}
    \Pi(\vec \mu_t) := \left\{ \sigma \in \cP(\mathbb{R}^d) \mid \sigma = {\rm Law} (X_t), \, {\rm Law}(X_{t,i}) = \mu_{t,i}, \; \forall i \in [d] \right\}.
  \end{align*}
  Clearly, if $\pi \in {\rm VMT}(\vec\mu)$, then $\pi^t \in \Pi(\vec \mu_t)$. 
  
  In the case $N=2$, we denote the marginals as $\vec \mu = (\mu_1, \dots, \mu_d)$ and $\vec \nu = (\nu_1, \dots, \nu_d)$, 
  where $\mu_i = {\rm Law}(X_i)$ and $\nu_i = {\rm Law}(Y_i)$ for $i \in [d]$. 
  
  Let $\mu, \nu \in \cP(\mathbb{R}^d)$ be two probability measures. We say they are in \emph{convex order}, denoted by $\mu \preceq_c \nu$, if and only if:
  \[
  \int f \, d\mu \leq \int f \, d\nu \quad \text{for every convex function } f.
  \]
  It is known that the set ${\rm VMT}(\vec\mu)$ is nonempty if and only if every pair of marginals $\mu_{t,i}, \mu_{t+1,i}$ is in \emph{convex order} \cite{kellerer1972}. Thus, we will always assume $\mu_{t,i} \preceq_c \mu_{t+1,i}$ for all $t \in [N-1]$ and $i \in [d]$.

  Let a measurable function $c : \mathbb{R}^{Nd} \to \mathbb{R}$ denote the cost or option payoff function. We define the vectorial martingale optimal transport (VMOT) problem as:
  \begin{align}\label{VMOT}
  \text{maximize} \quad \mathbb{E}_\pi [c(X)] \quad \text{over} \quad \pi \in {\rm VMT}(\vec\mu).
  \end{align}
  A solution $\pi$ to \eqref{VMOT} will be called a vectorial martingale optimal transport, or VMOT.
  
  To ensure that the problem \eqref{VMOT} is well-defined, we make the following assumptions throughout the paper. When considering a VMOT problem with a cost function $c$, we assume that the marginals satisfy the following condition: there exist continuous functions $v_{t,i} \in L^1(\mu_{t,i})$, $t \in [N]$ and $i \in [d]$, such that:
  \[
  |c(x_1, \dots, x_N)| \leq \sum_{t=1}^N \sum_{i=1}^d v_{t,i}(x_{t,i}).
  \]
  Note that this ensures:
  \[
  \left| \mathbb{E}_\pi [c(X)] \right| \leq \sum_t \sum_i \mathbb{E}_{\mu_{t,i}} [v_{t,i}(X_{t,i})] < \infty \quad \text{for any} \quad \pi \in {\rm VMT}(\vec\mu).
  \]
  It is standard to show that this in turn implies that the problem \eqref{VMOT} admits an optimizer whenever $c$ is upper-semicontinuous \cite{Villani09}.
 
  The conditions of the VMOT problem are inspired by \cite{bhp, DolinskySoner14, GaHeTo11, HenryLabordere, Hobson98, Hobson11}, 
  which originate from a closely related problem in mathematical finance: 
  the martingale optimal transport (MOT) problem. 
  Instead of considering the coupling of a vector of martingales from given one-dimensional marginals, 
  the MOT problem considers the coupling of a single martingale with a given marginal distribution, 
  which may be multi-dimensional.

  From a financial perspective, each sequence of random variables $(X_{1,i}, \dots, X_{N,i})$ represents an asset price process at $N$ future times $0 < t_1 < \cdots < t_N$. 
  Assuming a zero interest rate, each martingale measure $\pi \in {\rm VMT}(\vec \mu)$ represents the risk-neutral probability under which $(X_1, \dots, X_N) \in \mathbb{R}^{Nd}$ is an $\mathbb{R}^d$-valued $N$-period martingale. 
  \cite{BreedenLitzenberger} demonstrated that such marginal distribution information can be obtained from market data, 
  providing theoretical support for the model-free martingale optimal transportation approach we consider in this paper.
  The cost function $c(x_{1,1}, \dots, x_{1,d}, \dots, x_{N,1}, \dots, x_{N,d})$ represents an option whose payoff is path-dependent on the asset prices $(X_1, \dots, X_N)$.
  
  The key difference between VMOT and classical multi-dimensional MOT lies in the treatment of joint distributions among different assets at each period. 
  While the joint distribution between different assets (dimensions) is given in multi-dimensional MOT, 
  VMOT does not assume a predetermined joint distribution. 
  As a result, there is no guarantee that a coupling of the marginals at one period satisfies the convex order condition relative to any feasible coupling at the next period, 
  and this must be enforced separately.

  Additionally, VMOT is more suited for modeling real-world financial markets. 
  In classical multi-dimensional martingale optimal transport, 
  properly recovering the joint distribution requires observing a set of rainbow options \cite{talponenViitasaari2014}. 
  However, rainbow options are typically traded in the over-the-counter (OTC) market, 
  making them harder to observe and prone to mispricing due to their lower liquidity. 
  On the other hand, VMOT relies only on European vanilla options, 
  which are actively traded, making them easier to observe and less prone to mispricing. 
  This makes the VMOT framework more aligned with real-world market dynamics and more robust to potential market frictions.

  With this in mind, we consider the set of all possible laws ${\rm VMT}(\vec\mu)$ given the marginal information $\vec\mu$. 
  The max/min value in \eqref{VMOT} can then be interpreted as the upper/lower arbitrage-free price bound for the option $c$ among all risk-neutral measures consistent with the market data. 
  We define \eqref{VMOT} as a maximization problem, though it can also be framed as a minimization problem by simply changing $c$ to $-c$.

  Beyond its financial relevance, VMOT exhibits rich geometric properties similar to classical OT. 
  We extend the observations in \cite{EcksteinGuoLimObloj21, Lim22} and find that for the class of supermodular payoff functions, when there are two underlying assets $(d = 2)$, 
  the maximum arbitrage-free price is attained when the joint distribution of the underlying assets at the first maturity time is co-monotone. 
  This finding suggests a single-factor market structure for extreme pricing scenarios.
  Moreover, the monotonicity structure in VMOT helps reduce computational complexity: 
  since the dependence structure of assets at the first time is known explicitly, 
  only the dependence structure at subsequent times, as well as the martingale coupling structure across time periods, need to be computed.

  This paper is organized as follows.
  Section \ref{sec:problem_Main_Theorem} formally defines the VMOT problem and presents our main results, along with a conjecture for extending the monotonicity structure to higher dimensions.
  Section \ref{sec:dual_and_proof} defines the dual formulation of the VMOT problem and provides the proof of the main theorem.
  Section \ref{sec:counterexamples_and_related_examples} presents a counterexample to the conjecture in higher dimensions, along with other related examples in the VMOT problem. We provide a counterexample to show that this dimension reduction is sharp and does not extend to $d \ge 3$. 
  Finally, Section \ref{Numerics} demonstrates how our main result helps reduce computational complexity for the VMOT problem.
  The Appendix provides miscellaneous proofs.

  \section{Main Theorem and Conjecture}\label{sec:problem_Main_Theorem}

  Before we state the main theorem, we introduce the concept of supermodularity and irreducibility, which will be crucial in our analysis.
  
  \begin{definition}\label{def:supermodular}
  
  For $a, b \in \mathbb{R}^d$, we define $a \lor b$ as the componentwise maximum of $a,b$ and define $a \land b$ as the componentwise minimum, so that $(a \lor b)_i = \max \{a_i,b_i\}$ and $(a \land b)_i = \min \{a_i,b_i\}$. Let $d \ge 2$, and  $\bt : \mathbb{R}^d \to \mathbb{R} \cup \{+\infty\}$ be a function. Then submodularity and supermodularity of $\bt$ are defined, for all $a,b \in \mathbb{R}^d$, as:
  \begin{align*}
    \bt(a) + \bt(b) &\ge \bt(a \lor b) + \bt(a \land b), \\
    \bt(a) + \bt(b) &\le \bt(a \lor b) + \bt(a \land b),
  \end{align*}
  respectively. A function is called strictly submodular or strictly supermodular if the above inequality is strict for all $a, b \in \mathbb{R}^d$ with $\{a,b\} \neq \{a \lor b, a \land b\}$.
  
  If $\bt$ is twice differentiable, then $\bt$ is supermodular if $\frac{\partial^2 \bt}{\partial x_i \partial x_j} \ge 0$ for all $i \neq j$, and strictly supermodular if $\frac{\partial^2 \bt}{\partial x_i \partial x_j} > 0$ for all $i \neq j$. Hence, for example, the function $x \mapsto \sum_{1 \le i < j \le d} x_i x_j$ is strictly supermodular.
  
  \end{definition}
  
  \begin{definition}
    \leavevmode
    \begin{enumerate}[label=\roman*), ref=\roman*]
      \item $\{ a \lor b, a \land b \}$ is called a monotone rearrangement of $\{a,b\}$.
      \item A set $A \subset \mathbb{R}^d$ is called monotone if for any $a,b \in A$, $\{a,b\} = \{a \lor b, a \land b\}$.
      \item A measure $\mu \in \cP(\mathbb{R}^d)$ is called monotone (or monotonically supported) if there is a monotone set $A$ such that $\mu$ is supported on $A$, i.e., $\mu(A) = 1$.
      \item Given a vector of probability measures $\vec \mu = (\mu_1, \dots, \mu_d)$ where each $\mu_i \in \cP(\mathbb{R})$, the unique probability measure $\chi_{\vec\mu} \in \cP(\mathbb{R}^d)$, which is monotone and has $\mu_1, \dots, \mu_d$ as its marginals (i.e., $\chi_{\vec\mu} \in \Pi(\vec\mu)$), is called the monotone coupling of $\vec \mu$. 
    \end{enumerate}
  \end{definition}
Note that $\chi_{\vec\mu} =(F^{-1}_{\mu_1}, F^{-1}_{\mu_2}, ..., F^{-1}_{\mu_d})_\#\mathcal{L}_{[0,1]}$,\footnote{Given a measurable map $F : \mathcal{X} \to \mathcal{Y}$ and a measure $\mu$ on $\mathcal{X}$, the push-forward of $\mu$ by $F$, denoted by $F_\# \mu$, is a measure on $\mathcal{Y}$ satisfying $F_\# \mu(A) = \mu(F^{-1}(A))$ for every $A \subset \mathcal{Y}$.} 
      where each $F^{-1}_{\mu_i}$ denotes the generalized inverse of the cumulative distribution function of $\mu_i$ (i.e., the quantile function), defined by $F^{-1}(y) = \inf \{x \in \mathbb{R} : F(x) \geq y\}$ for $y \in [0, 1]$, and $\mathcal{L}_{[0,1]}$ denotes the uniform probability measure on $[0,1]$.
      
  We now discuss \emph{irreducibility} between two probability measures. 
  A pair of measures $(\mu, \nu)$ on $\R$ in convex order is said to be \emph{irreducible} if the set $I := \{ x \in \mathbb{R} \mid \int |x-y| \, d(\nu-\mu)(y) > 0 \}$ is connected and contains the full mass of $\mu$.
  We say that a sequence of measures $(\mu_t)_{t \in [N]}$ is \emph{irreducible} if each consecutive pair $(\mu_t, \mu_{t+1})$ is irreducible.

  We note that the irreducibility condition imposed on the sequence of marginals holds for many pairs of probability distributions $\mu_t \preceq_c \mu_{t+1}$ on the real line in convex order. 
  Moreover, if a pair of measures is not irreducible, it can be perturbed in an arbitrarily small way to make the pair irreducible. 
  For further details on irreducibility, we refer the reader to \cite{bnt}. 
  
\subsection{Main Theorem}

We first present the main theorem of this paper, 
which establishes the existence of a VMOT with specific properties for the first-period marginals. 

\begin{theorem}\label{main}
  Let $d = 2$ and $N \geq 2$. Consider the cost function $c(x_1, \dots, x_N) = \sum_{t=1}^N c_t(x_{t,1}, x_{t,2})$, 
  where each $c_t$ is a supermodular function. 
  Assume the following conditions:
  \begin{enumerate}[label=\roman*), ref=\roman*]
    \item For each $i =1,2$, $(\mu_{t,i})_{t \in [N]}$ is an irreducible sequence of marginals on $\R$.
    \item The cost function $c: \R^{2N} \to \R$ is upper-semicontinuous, and there exist continuous functions $v_{t,i} \in L^1(\mu_{t,i})$ such that $|c(x)| \leq \sum_{t=1}^N \sum_{i=1}^2 v_{t,i}(x_{t,i})$ for all $x \in \mathbb{R}^{2N}$.
    \item The second moments of $\mu_{1,1}$ and $\mu_{1,2}$ are finite.
  \end{enumerate}
  Then:
  \begin{enumerate}[label=\roman*), ref=\roman*]
    \item There exists a VMOT $\pi$ such that its first period marginal $\pi^1$ is the monotone coupling of $\mu_{1,1}$ and $\mu_{1,2}$.
    \item If $c_1$ is strictly supermodular, then every VMOT $\pi$ satisfies that its first period marginal $\pi^1$ is the monotone coupling of $\mu_{1,1}$ and $\mu_{1,2}$.
  \end{enumerate}
\end{theorem}

Using the proof of Theorem \ref{main} and Proposition \ref{prop1} iii) in Section \ref{sec:proof}, we also obtain the following mirror statement. A set $A \subset \mathbb{R}^2$ is called {\em anti-monotone} if the set $\{ x = (x_1, x_2) \in \mathbb{R}^2 \mid (-x_1, x_2) \in A \}$ is monotone. A measure $\mu \in \cP(\mathbb{R}^2)$ is called anti-monotone if $\mu$ is supported on an anti-monotone set.

\begin{corollary}\label{mainCorollary}
  Let $d = 2, N \geq 2$ and $c(x_1, \dots, x_N) = \sum_{t=1}^N c_t(x_{t,1}, x_{t,2})$, where each $c_t$ is submodular. 
  Assume the same conditions as in Theorem \ref{main}. Then:
  
  \begin{enumerate}[label=\roman*), ref=\roman*]
    \item There exists a VMOT $\pi$ such that its first period marginal $\pi^1$ is the anti-monotone coupling of $\mu_1$ and $\mu_2$.
    \item If $c_1$ is strictly submodular, then every VMOT $\pi$ satisfies that its first period marginal $\pi^1$ is the anti-monotone coupling of $\mu_{1,1}$ and $\mu_{1,2}$.
  \end{enumerate}
\end{corollary}

\begin{remark}
  The (anti-)monotonicity of the first period implies that there exists a random variable $Z$ such that $X_{1,1} = f_1(Z)$ and $X_{1,2} = f_2(Z)$, where $f_1, f_2$ are deterministic functions. 
  A financial interpretation may be given as follows: for an $N$-period cap\footnote{Option payoffs of the form $c(x_1,\dots,x_N) = \sum_{t=1}^N c_t(x_t)$ are sometimes referred to as an $N$-period cap, where $x_t=(x_{t,1},...,x_{t,d})$ denotes the prices of $d$ assets at the $t^{\rm th}$ maturity.} 
  whose payoff depends on two underlying assets, Theorem \ref{main} and Corollary \ref{mainCorollary} describe the existence of an extreme market model that attains the extremal of the price bounds. Specifically, this market model implies that the two assets at the first period $(X_{1,1}, X_{1,2})$ are controlled by a single factor $Z$. Furthermore, if $c_1$ is strictly submodular or supermodular, then every extremal market model exhibits this property.
\end{remark}

\subsection{The Conjecture}

Theorem \ref{main} establishes that VMOT exhibits a monotone structure for $d = 2$ and $N \geq 2$ when the cost function is supermodular. 
This result highlights the critical role of supermodular costs in ensuring monotone structural properties within VMOT, 
a concept analogous to its well-established role in classical multi-marginal optimal transport problems. 

This raises a natural question: can these structural properties extend to VMOT in higher dimensions, 
specifically for $d \ge 3$? We propose the following conjecture:

\begin{conjecture}\label{conjecture1}
  Let $d \geq 2$, $N \geq 2$, and assume the following conditions:
  
  \begin{enumerate}[label=\roman*), ref=\roman*]
    \item The sequence of marginals $(\mu_{t,i})_{t \in [N]}$ is irreducible for each $i \in [d]$, and in convex ordering, i.e., $\mu_{t,i} \preceq_c \mu_{t+1,i}$ for all $t \in [N-1]$ and $i \in [d]$.
    \item The cost function $c(x_1, \dots, x_N) = \sum_{t=1}^N c_t(x_{t,1}, \dots, x_{t,d})$ is upper semicontinuous with supermodular $c_t$ for all $t \in [N]$, and $c$ is absolutely bounded by the sum of continuous functions $v_{t,i}(x_{t,i}) \in L^1(\mu_{t,i})$ for each $t \in [N]$ and $i \in [d]$.
    \item The second moments of $\mu_{1,i}$ are finite for all $i \in [d]$.
  \end{enumerate}
  
  Then:
  
  \begin{enumerate}[label=\roman*), ref=\roman*]
    \item There exists a VMOT $\pi$ for the problem \eqref{VMOT} such that its first period marginal $\pi^1$ is the monotone coupling of $\mu_1 = (\mu_{1,1}, \dots, \mu_{1,d})$. 
    \item If $c_1$ is strictly supermodular, then every VMOT $\pi$ satisfies that its first period marginal $\pi^1 = \chi_{\mu_1}$, where $\chi_{\mu_1}$ is the monotone coupling of $\mu_1$.
  \end{enumerate}
\end{conjecture}
     
This conjecture is based on the intuition that supermodularity may induce monotonicity in the first period. However, the martingale constraints in subsequent periods reduce the likelihood that monotonicity will be maintained across all periods. Theorem \ref{main} confirms the validity of the conjecture for two assets with an arbitrary number of periods $N$.

\subsection{Motivation and Heuristic}

To understand both the main theorem and the conjecture, 
we draw motivation from the classical multi-dimensional optimal transport setting. 
In this framework, the monotonicity of the optimal measure is guaranteed for supermodular costs, 
regardless of the dimension $d$:

\begin{theorem}\label{fact1}\cite{Carlier03, Lorentz53}
  If $c$ is supermodular, 
  the monotone coupling $\chi_{\vec\mu}$ arises as a maximizer of $\E_\ga[c(X)] = \int_{\R^d} c(x)\, d\ga(x)$ among all $\ga \in \Pi(\vec \mu)$, 
  and $\chi_{\vec\mu}$ is the unique maximizer if $c$ is strictly supermodular.
\end{theorem}

The monotonicity of optimizers in the classical optimal transport problem for supermodular costs is well-known \cite{Carlier03, Lorentz53}. 
Results asserting higher-dimensional deterministic solutions, such as those by Brenier \cite{Brenier87} (for two marginals) and Gangbo-{\'S}wi{\k{e}}ch \cite{GangboSwiech98} (for three or more marginals) regarding the cost function $c(x) = \sum_{1 \le i < j \le d} x_i \cdot x_j$, 
and generalizations to other cost functions \cite{Levin99, GangboMcCann96, Caffarelli96} (for two marginals), \cite{Heinich2002, Pass2011, KimPass2014, PassVargasJimenez2021} (for several marginals), can be seen as higher-dimensional analogues of this monotonicity. Our conjecture can be interpreted as a vectorial martingale transport version of these results. Indeed, note that if $N=2$ and each $\mu_i$ is a Dirac mass (corresponding to the case where the first time is the present), the VMOT problem reduces to the classical (multi-marginal) optimal transport problem on the $\nu_i$'s.

\noin {\bf Heuristic.} Given marginals $\vec \mu = (\vec\mu_1, \dots, \vec\mu_N)$ such that $\vec\mu_t = (\mu_{t,1},...,\mu_{t,d})$ for $t \in [N]$ 
and a cost function $c(x_1,\dots,x_N) = \sum_{t=1}^N c_t(x_t)$ where each $c_t$ is a supermodular function, 
in light of Theorem \ref{fact1}, the ideal situation would be that for the marginals $\pi^t$ of each period of $\pi \in {\rm VMT}(\vec \mu)$, 
we have the monotone coupling $\pi^t = \chi_{\vec \mu_t}$, such that $\E_\pi [c(X)] = \sum_{t=1}^N \E_{\chi_{\vec \mu_t}} [c_t(X)]$. However, the martingale constraint imposed on $\pi$ reduces the possibility of such an ideal situation. 

Even if $\mu_{t,i} \preceq_c \mu_{t+1,i}$ for all $i \in [d]$, the monotone couplings $\chi_{\vec \mu_t}$ and $\chi_{\vec \mu_{t+1}}$ may not satisfy the convex order in general. 
As an example, consider the following case: 
\begin{example}
Let $d = 2$, $\mu_1 = \mu_2$ be the uniform probability measure on the interval $[-1,1]$, $\nu_1$ be uniform on $[-3,3]$, and $\nu_2$ be uniform on $[-2,2]$. 
Then $\chi_{\vec \mu}$ is the uniform probability on $l_1 = \{(x_1, x_2) \in \mathbb{R}^2 \mid x_1 = x_2, \, x_1 \in [-1,1]\}$, 
and $\chi_{\vec \nu}$ is uniform on $l_2 = \{(y_1, y_2) \in \mathbb{R}^2 \mid y_2 = \frac{2}{3} y_1, \, y_1 \in [-3,3]\}$. 
Then $\chi_{\vec \mu}$ and $\chi_{\vec \nu}$ cannot be in convex order because it is necessary for $l_1$ to be contained in the convex hull of $l_2$ to satisfy $\chi_{\vec \mu} \preceq_c \chi_{\vec \nu}$.
\end{example}

\begin{figure}[h!]
  \centering
  \subfloat[Generic marginals of a two-period vectorial martingale transport.\label{badmart}]{%
    \includegraphics[width=0.89\textwidth]{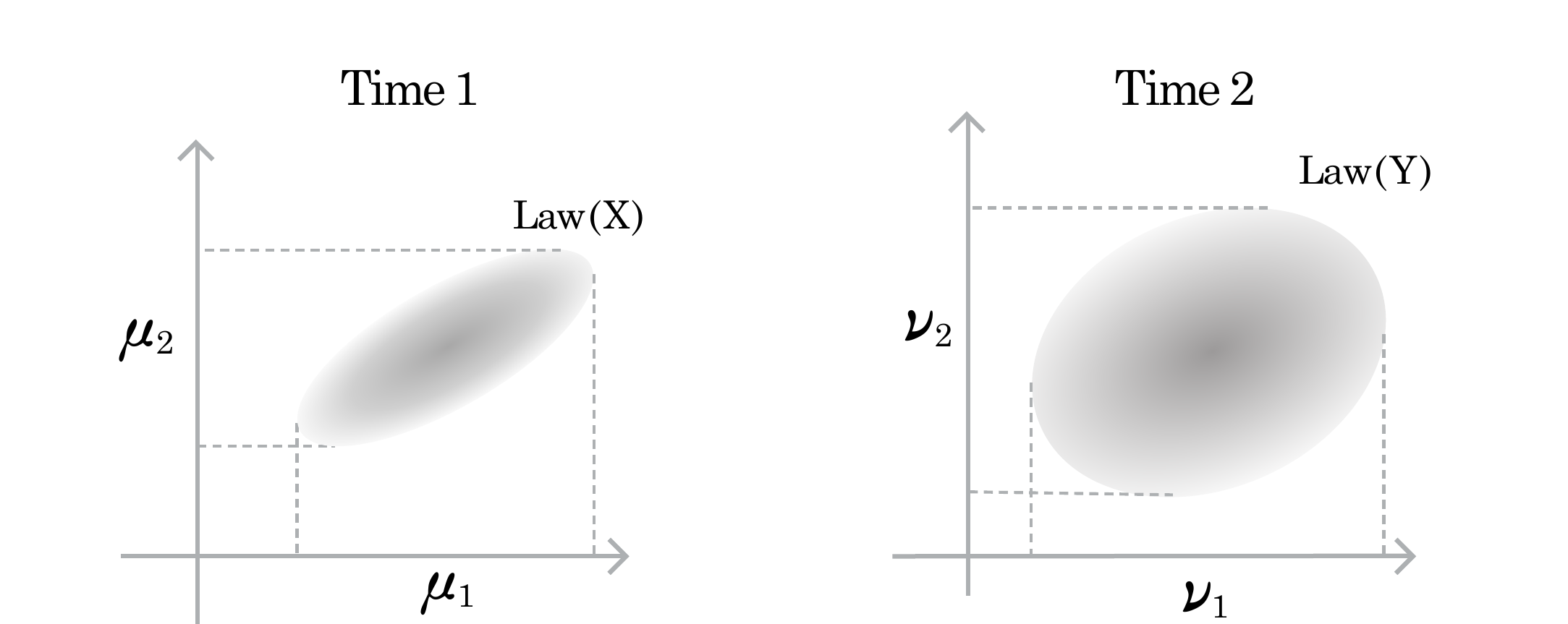}%
  }
  \par 
  \subfloat[A martingale transport with a monotone first-time marginal.\label{goodmart}]{%
    \includegraphics[width=0.89\textwidth]{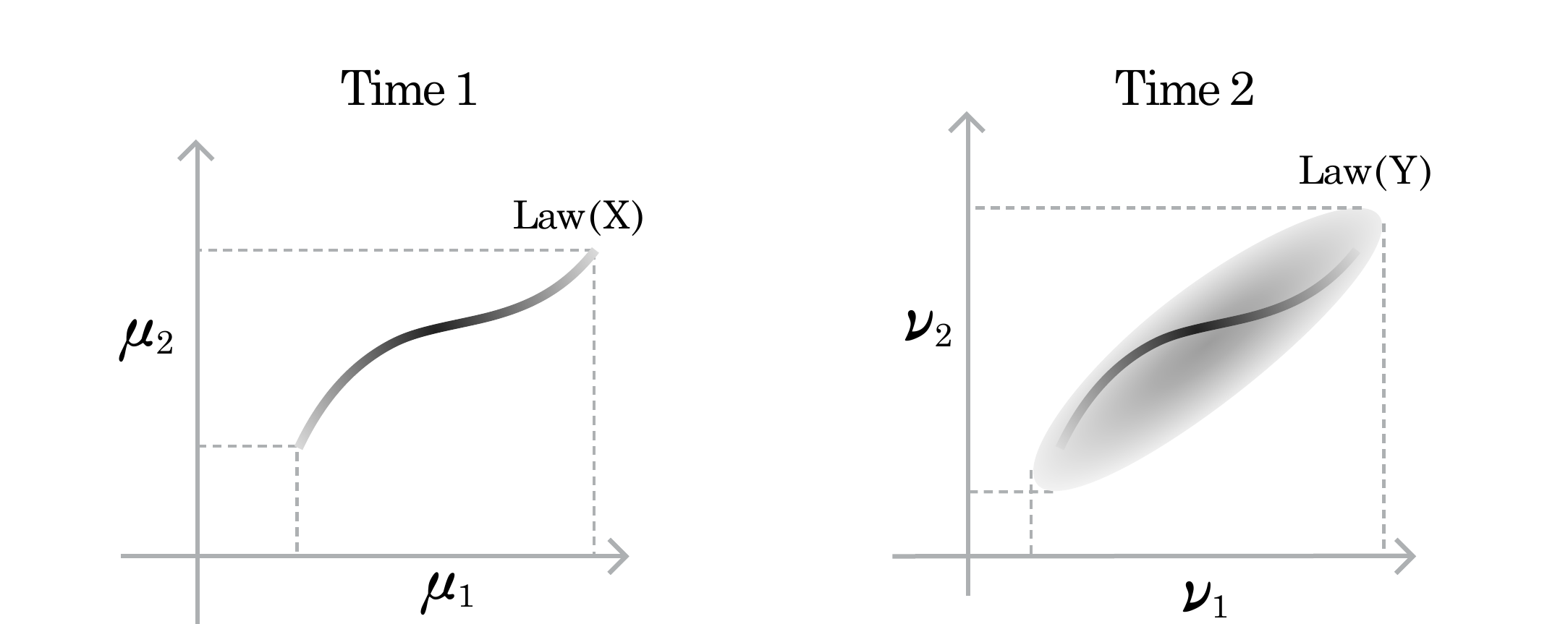}%
  }
  \caption{Conjecture \ref{conjecture1} asserts that the martingale transport in Figure \ref{goodmart} can be superior to the one in Figure \ref{badmart} for the problem \eqref{VMOT} due to the supermodularity of $c_1, c_2$. We overlap ${\rm Law}(X)$ with ${\rm Law}(Y)$ in Figure \ref{goodmart} to emphasize that they must be in convex order.}
  \label{mart}
\end{figure}

The intuition behind this is that the convex order condition in each of the marginals is not sufficient to guarantee that the convex hull of the monotone set induced by the second-period marginals contains the first-period monotone set. 
This demonstrates that the ideal situation is infeasible in general. Nevertheless, it is plausible that a VMOT $\pi$ can still couple the marginals $\vec \mu_1$ monotonically. 
\footnote{For any $\ga \in \Pi(\vec \mu)$, there exists $\pi \in {\rm VMT}(\vec \mu, \vec \nu)$ such that $\pi^1 = \ga$. 
One can simply take the disintegration kernels $\pi^{1,i}$ of some $\pi_i \in \Pi^M(\mu_i, \nu_i)$, where $\Pi^M(\mu_i, \nu_i)$ denotes the set of all martingale couplings, and set $\pi = \ga(dx_1, \dots, dx_d) \otimes \pi^{1,1}(dy_1) \otimes \dots \otimes \pi^{1,d}(dy_d)$. 
Similar arguments can be made for general $N$.}
This leads us to our heuristic behind the conjecture, 
as illustrated in Figure \ref{mart}. 

\begin{remark}
  \cite{EcksteinGuoLimObloj21} showed that the conjecture holds for $N=2$ and general $d$ when the cost function is of the quadratic form \eqref{portfolio_cost}, 
  and the marginals $\mu_i, \nu_i$ satisfy a special condition known as the linear increment of marginals. 
  This is the case when, for example, $\mu_i, \nu_i$ are Gaussians with increasing variance. 
  In fact, \cite{EcksteinGuoLimObloj21} showed that under these conditions, $\pi^1$ of a VMOT $\pi$ is distributed on a straight line in $\mathbb{R}^d$. 
  This implies that the assumptions on the cost and marginals are very restrictive, 
  and the conclusion cannot be extended to general marginals. See \cite{Lim22}
  for a related discussion. 
  In this paper, we will prove that the conjecture is correct when $d=2$ (without any particular condition on the marginals) but incorrect in general when $d \geq 3$. 
  It is unfortunate that we cannot find a general sufficient condition for the marginals $\mu_{t,i}$ other than the condition given by \cite{EcksteinGuoLimObloj21} to ensure the conjecture holds for general $d$.
\end{remark}

\subsection{Examples of Supermodular Costs}

To illustrate the importance of supermodular costs in finance, 
we consider the following cost function:
\be\label{mutualcost}
c(x,y) = \sum_{1 \le i , j \le d} (a_{ij}x_ix_j + b_{ij}x_iy_j + c_{ij}y_iy_j),
\ee
which represents the assets' mutual covariances at two future times. 
For any $\pi \in {\rm VMT}(\vec\mu,\vec\nu)$, $\E_\pi [ X_i Y_j] = \E_\pi[\E_\pi[ X_i Y_j|X]] = \E_\pi[X_i\E_\pi[ Y_j|X]] = \E_\pi[X_i X_j]$ by the martingale constraint, 
while $\E_\pi [X_i^2] = \int_\R x^2 d\mu_i(x)$, $\E_\pi [Y_j^2] = \int_\R y^2 d\nu_j(y)$ are fixed by the marginal constraints. 
Hence, we can reduce the cost \eqref{mutualcost} to the following form:
\be\label{cost}
c(x,y) = \sum_{1 \le i < j \le d} (a_{ij}x_ix_j + b_{ij}y_iy_j).
\ee
We assume $a_{ij} \ge 0$, $b_{ij} \ge 0$ to ensure that the cost function is supermodular.

In particular, for $d=2$, this becomes
\[
c(x,y) = a x_1x_2 + b y_1y_2,
\]
so that $\E_\pi[c] = a\E_\pi[X_1 X_2] + b\E_\pi[Y_1 Y_2]$ represents a weighted sum of mutual covariances between $X_1, X_2$ and between $Y_1, Y_2$ under the market model $\pi$. 

To motivate the study of the cost function of the form \eqref{cost}, 
consider a portfolio consisting of assets with prices $Y_1, \dots, Y_d$ at the terminal maturity $t_2$, 
with weights $w_1, \dots, w_d > 0$, so that the value of the portfolio at maturity is $\sum_{i=1}^d w_i Y_i$. 
The variance is a commonly used measure of the portfolio's risk:
\[
{\rm Var}_\pi\left(\sum_{i=1}^d w_iY_i\right) = \mathbb{E}_{\pi}\left[\left(\sum_{i=1}^d w_i Y_i\right)^2\right] - \left(\sum_{i=1}^d w_i \mathbb{E}_{\pi}[Y_i]\right)^2, \quad \text{given } \pi \in {\rm VMT}(\vec\mu, \vec\nu).
\]

In the VMOT framework, the term $\left(\sum_{i=1}^d w_i \mathbb{E}[Y_i]\right)^2$ does not depend on $\pi$, 
since the individual distributions are assumed to be known. 
Maximizing the variance is thus equivalent to maximizing $\mathbb{E}_{\pi^Y} \left[\left(\sum_i w_i Y_i\right)^2\right] = \int_{\mathbb{R}^d} \left(\sum_i w_i y_i\right)^2 d\pi^Y(y)$, 
which is a classical (multi-marginal) optimal transport (OT) problem with cost function:
\be\label{portfolio_cost}
c(x,y) = c(y) = \left(\sum_{i=1}^d w_i y_i\right)^2.
\ee

Notice that the optimization problems with $c$ as in \eqref{cost} and $c$ as in \eqref{portfolio_cost} are equivalent, 
with each $a_{ij} = 0$ and $b_{ij} = w_i w_j$ in the VMOT problem, 
again due to the given marginal information.

Similar to OT, the VMOT problem reflects a worst-case risk scenario in which, 
in addition to the distributions $\nu_i$ at some future time $t_2$, 
the distributions $\mu_i$ of the assets at an earlier time $t_1 < t_2$ are known. 
In this situation, one may still try to evaluate the risk or variance of the portfolio's value $\sum_{i=1}^d w_i Y_i$ at time $t_2$, 
incorporating the extra information from the knowledge of distributions $\mu_i$ at $t_1$. 
This leads to the VMOT problem with cost \eqref{portfolio_cost}. 
As a result, the VMOT approach can yield tighter bounds on the portfolio's variance than the classical OT approach, 
even though the cost functions depend only on the terminal prices $Y$. The reason is that only couplings $\pi^2$ of the marginals $\nu_i$ which dominate some coupling $\pi^1$ of the marginals $\mu_i$ in convex order can arise as feasible couplings in VMOT, 
whereas all couplings in $\Pi(\vec \nu)$ are allowed as candidate optimizers in the classical OT.

It is worth noting that the maximal implied variance, which is the solution of the VMOT problem when the marginals are the risk-neutral distributions of the $Y_i$, 
differs from the physical variance, which is derived by the physical joint distributions of the $Y_i$. 
Importantly, option prices can be used to calculate the risk-neutral variance, which is widely used as a proxy for true variance. 
Empirical evidence suggests that option-implied ex-ante higher moments hold predictive power for future stock returns \cite{ConardDittmarGhysels13}. 
This indicates that option prices encompass market information and reflect investors' expectations of future stock moments \cite{JackwerthRubinstein96}. 
The VMOT approach, which derives marginal distributions from option prices, resonates with these findings.

Table \ref{submodularPayoff} concludes this section by presenting several supermodular payoff functions for exotic options, 
some of which are also found in \cite{AnsariLutkebohmertNeufeldSester22}. 
The $N$-period cap with the payoff $c(x_1, \dots, x_N) = \sum_{t=1}^N c_t(x_t)$, 
where $c_t$ is one of the supermodular functions listed in Table \ref{submodularPayoff}, 
written on two underlying assets, fits into the form of the cost function applicable to our Theorem \ref{main}.

\begin{table}[h]
\centering
\begin{tabular}{|l|c|}
\hline
Name                                                & Payoff function                                           \\ \hline
European basket call option                         & $(\sum_{i = 1}^d a_{i} X_{i} - K)^+$                \\ \hline
European basket put option                          & $(K - \sum_{i = 1}^d a_i X_{i})^+$                    \\ \hline
Put on the maximum among $d$ stocks                 & $(K - \max_{1 \leq i \leq d}\{X_{i}\})^+$               \\ \hline
Call on the minimum among $d$ stocks                & $(\min_{1 \leq i \leq d}\{X_{i}\} - K)^+$               \\ \hline
Covariance among $d$ stocks                         & $\sum_{i,j} a_{ij}X_{i}X_{j}$               \\ \hline
\end{tabular}
\vspace{3mm}

\caption{\label{submodularPayoff} A list of supermodular derivative payouts with nonnegative coefficients $a_{ij}\ge 0$.}
\end{table}

\section{Dual Problem, Dual Attainment and Proof of the Main Theorem}\label{sec:dual_and_proof}

\subsection{Dual Problem}

The VMOT problem is a type of infinite-dimensional linear programming, 
which admits a corresponding {\em dual problem}. We introduce the space of functions 
\begin{align*}
\mathcal{H} = \bigg\{ \varphi(x) &\coloneqq \sum_{t=1}^N \sum_{i=1}^d u_{t,i}(x_{t,i}) 
+ \sum_{t=1}^{N-1} \sum_{i=1}^d h_{t,i}(x_1, \dots, x_t) (x_{t+1,i} - x_{t,i}) \\
&\quad \bigg|\, u_{t,i} \in L^1(\mu_{t,i}), \ \  t \in [N], \, i \in [d]; \quad h_{t,i} \in L^\infty(\mathbb{R}^{td}), \ \ t \in [N-1], \, i \in [d] \bigg\}.
\end{align*}

The space \( \mathcal{H} \) consists of the Lagrange multipliers associated with the marginal and martingale constraints in the dual problem. 

We then define the set \( \Xi \) as the subset of \( \mathcal{H} \) satisfying the pointwise inequality:
\be\label{ptwiseineq}
\Xi := \big\{ \varphi \in \mathcal{H} \mid \varphi(x) \geq c(x) \text{ for all } x \in \mathbb{R}^{Nd} \big\}.
\ee
For any martingale measure $\pi \in {\rm VMT}(\vec \mu)$ and $\varphi \in \cH$, it holds that $\int \varphi \, d\pi = \sum_{t,i}\int u_{t,i}d\mu_{t,i}$. This leads to the dual problem associated with \eqref{VMOT} as: 
\begin{equation}\label{dualproblem}
  \inf_{\varphi \in \Xi} \sum_{t=1}^N \sum_{i=1}^d \int u_{t,i} \, d\mu_{t,i}.
\end{equation}

In this formulation, \eqref{VMOT} can be referred to as the {\em primal problem}. 
In the special case of $N=2$, we denote $\Xi$ as the set of functions of the form $(u_{1,i}, u_{2,i}, h_{1,i})_{i=1}^d = (u_i, v_i, h_i)_{i=1}^d$.

The dual problem has a concrete financial interpretation. 
Specifically, the inequality in \eqref{ptwiseineq} represents a semi-static super-replicating portfolio for the path-dependent payoff $c$. 
Here, $u_{t,i}(x_{t,i})$ corresponds to the European options written on each individual underlying asset, 
with maturities at time $t$, and $h_{t,i}$ represents the position held in the $i^\textrm{th}$ asset for delta-hedging between times $t$ and $t+1$. 

If the problem \eqref{VMOT} is a minimization problem, 
then the dual problem corresponds to an optimal subhedging problem, 
where the inequality in \eqref{ptwiseineq} is reversed. 

The value in \eqref{dualproblem} represents the minimum cost required to construct a superhedging portfolio. Thus, the goal of the dual problem is to identify the optimal (least-cost) superhedging portfolio.

\subsection{Dual Attainment}

It has been established that the dual problem \eqref{dualproblem} generally does not admit a solution within the class $\Xi$, 
even in the simple case $d=1$ and $N=2$, i.e., when the option $c$ depends on a single asset (see \cite{bj, bnt}). 
This issue arises unless certain regularity conditions are imposed on the payoff function $c$ \cite{blo}. 
Consequently, a generalized notion of {\em dual attainment}, which refers to the solvability of the dual problem, 
has been introduced for multi-period problems. 
This was studied in \cite{NutzStebeggTan20} for $N \geq 2, d = 1$, 
and in \cite{Lim23} for $N \geq 2, d \geq 2$. 
More specifically, \cite{Lim23} provides the following result on dual attainment.

\begin{theorem}[\cite{Lim23}]\label{main2}
  Let $(\mu_{t,i})_{t \in [N]}$ be an irreducible sequence of marginals on $\R$ for each $i \in [d]$.
  Let $c: \R^{Nd} \to \R$ be an upper-semicontinuous cost such that $|c(x)| \le \sum_{t=1}^N \sum_{i=1}^d v_{t,i}(x_{t,i})$ for some continuous functions $v_{t,i} \in L^1(\mu_{t,i})$. Then there exists a {\em dual minimizer}, consisting of two sets of functions $(u_{t,i})_{t \in [N], i \in [d]}$ and $(h_{t,i})_{t \in [N-1], i \in [d]}$, that satisfies the following pathwise constraint:
  \be\label{ptwiseeq}
    \sum_{t=1}^N \sum_{i=1}^d u_{t,i}(x_{t,i}) + \sum_{t=1}^{N-1} \sum_{i=1}^d h_{t,i}(x_1, \dots, x_t) \left( x_{t+1,i} - x_{t,i} \right) = c(x) \q  \pi-a.e. \, x
  \ee
  for every {\rm VMOT} $\pi$ that solves the primal problem \eqref{VMOT}.
\end{theorem}

\subsection{Proof of the Main Theorem}\label{sec:proof}
Before proceeding with the proof of the main theorem, 
we first present some useful properties of supermodular functions.

\begin{definition}
For a proper function $f : \R^d \to \R \cup \{+\infty\}$, 
its convex conjugate $f^*$ is defined as the following convex lower semi-continuous function:
\begin{equation}
f^*(y) = \sup_{x \in \R^d} \left( x \cdot y - f(x) \right), \quad y \in \R^d.
\end{equation}
\end{definition}

It is well-known that $f^{**} = (f^*)^*$ is the largest convex lower semi-continuous function satisfying $f^{**} \le f$. 
We refer to $f^{**}$ as the {\em convex biconjugate} of $f$.

\begin{proposition}\label{prop1}
  \leavevmode
  \begin{enumerate}[label=\roman*), ref=\roman*]
    \item If $\beta$ is a submodular function on $\mathbb{R}^d$, then its convex conjugate $\beta^*$ is supermodular.
    \item If $\beta$ is supermodular on $\mathbb{R}^2$, then its convex conjugate $\beta^*$ is submodular.
    \item If $\beta$ is submodular or supermodular on $\mathbb{R}^2$, then its convex biconjugate $\beta^{**}$ is also submodular or supermodular, respectively.
  \end{enumerate}
\end{proposition}

The proof of Proposition \ref{prop1} is provided in the appendix.

With the dual attainment theorem and the proposition about the convex biconjugate of submodular and supermodular functions, 
we now proceed with the proof of Theorem \ref{main}.

\begin{proof}[of Theorem \ref{main}]
  Theorem \ref{main2} guarantees the existence of an optimal dual solution $(u_{t,i})_{t \in [N],i \in [d]}$ and $(h_{t,i})_{t \in [N-1], i \in [d]}$ that satisfies:
  \[
    \sum_{t=1}^N \sum_{i=1}^d u_{t,i}(x_{t,i}) + \sum_{t=1}^{N-1} \sum_{i=1}^d h_{t,i}(x_1, \dots, x_t) \left( x_{t+1,i} - x_{t,i} \right) \geq \sum_{t=1}^N c_t(x_t),
  \]
  for all $x \in \R^{Nd}$, with equality $\pi$-almost every $x$ for any VMOT $\pi$.

  Rearranging terms, we obtain a linear function of $x_N$ on the right-hand side:
  \[
    \sum_{i=1}^d u_{N,i}(x_{N,i}) - c_N(x_N) \geq \sum_{t=1}^{N-1} \left( c_t(x_t) - \sum_{i=1}^d u_{t,i}(x_{t,i}) \right) - \sum_{t=1}^{N-1} \sum_{i=1}^d h_{t,i}(x_1, \dots, x_t) \left( x_{t+1,i} - x_{t,i} \right),
  \]
  for all $x \in \R^{Nd}$, with equality $\pi$-almost every $x$. Let
  \[
    \bt_N(x_N) := \sum_{i=1}^d u_{N,i}(x_{N,i}) - c_N(x_N).
  \]
  Since the convex biconjugate of a function is the supremum of all affine functions below it, we have
  \be\label{ptwiseineq_2}
    \bt_N^{**}(x_N) \geq \sum_{t=1}^{N-1} \left( c_t(x_t) - \sum_{i=1}^d u_{t,i}(x_{t,i}) \right) - \sum_{t=1}^{N-1} \sum_{i=1}^d h_{t,i}(x_1, \dots, x_t) \left( x_{t+1,i} - x_{t,i} \right),
  \ee
  for all $x \in \R^{Nd}$, with equality $\pi$-almost every $x$. In particular, by letting $x_N = x_{N-1}$, we obtain
  \[
    \beta_N^{**}(x_{N-1}) \geq \sum_{t=1}^{N-1} \left( c_t(x_t) - \sum_{i=1}^d u_{t,i}(x_{t,i}) \right) - \sum_{t=1}^{N-2} \sum_{i=1}^d h_{t,i}(x_1, \dots, x_t) \left( x_{t+1,i} - x_{t,i} \right),
  \]
  for all $(x_1, \dots, x_{N-1}) \in \R^{(N-1)d}$.

  Next, disintegrating any VMOT $\pi$ with respect to $x_1, \dots, x_{N-1}$, thereby  representing\\ 
  $\pi(dx_1, \dots, dx_N) = \kappa(x_1, \dots, x_{N-1}, dx_N) \otimes \pi^{1, \dots, N-1}(dx_1, \dots, dx_{N-1})$, we integrate \eqref{ptwiseineq_2} with respect to the martingale kernel $\kappa(x_1, \dots, x_{N-1}, dx_N)$, and obtain
  \begin{align*}
    \sum_{t=1}^{N-1} \left( c_t(x_t) - \sum_{i=1}^d u_{t,i}(x_{t,i}) \right) - &\sum_{t=1}^{N-2} \sum_{i=1}^d h_{t,i}(x_1, \dots, x_t) \left( x_{t+1,i} - x_{t,i} \right)\\
    &= \int \bt_N^{**}(x_N) \kappa(x_1, \dots, x_{N-1}, dx_N) \\
    &\geq \bt_N^{**}(x_{N-1}) \quad \pi^{1, \dots, N-1} \text{-a.e. } (x_1, \dots, x_{N-1}),
  \end{align*}
  since $\int h_{N-1,i}(x_1, \dots, x_{N-1}) \left( x_{N,i} - x_{N-1,i} \right) \kappa(x_1, \dots, x_{N-1}, dx_N) = 0$ by the martingale property, while the last inequality is from Jensen's inequality for the convex function $\bt_N^{**}$. 
  We conclude:
  \[
    \beta_N^{**}(x_{N-1}) \geq \sum_{t=1}^{N-1} \left( c_t(x_t) - \sum_{i=1}^d u_{t,i}(x_{t,i}) \right) - \sum_{t=1}^{N-2} \sum_{i=1}^d h_{t,i}(x_1, \dots, x_t) \left( x_{t+1,i} - x_{t,i} \right)
  \]
  for all $(x_1, \dots, x_{N-1}) \in \R^{(N-1)d}$, with equality $\pi^{1, \dots, N-1}$-almost every $(x_1, \dots, x_{N-1})$.

  By inductively defining the following function  
  \[
    \bt_t(x_t) := \bt_{t+1}^{**}(x_t) + \sum_{i=1}^d u_{t,i}(x_{t,i}) - c_t(x_t),
  \]
for $t=N-1, N-2, \dots, 2$ and continuing this process, we deduce
  \[
    \beta_2^{**}(x_1) - c_1(x_1) \geq -\sum_{i=1}^d u_{1,i}(x_{1,i}),
  \]
  for all $x_1 \in \mathbb{R}^d$, with equality $\pi^1$-almost every $x_1$.

  This implies that $\pi^1$ solves the optimal transport problem with cost $\bt_2^{**}(x_1) - c_1(x_1)$ and marginals $\mu_{11}, \dots, \mu_{1d}$.

  As each $c_t$ is supermodular, $\bt_N$ is submodular. Assuming $d = 2$ and $c_1$ is strictly supermodular, by Proposition \ref{prop1} iii), each $\beta_t^{**}$ is submodular, making $\beta_2^{**}(x_1) - c_1(x_1)$ strictly submodular. By Theorem \ref{fact1}, $\pi^1 = \chi_{(\mu_{11}, \mu_{12})}$ must be the monotone coupling of $\mu_{11}, \mu_{12}$, proving part ii).

  To prove part i), fix $\delta > 0$ and choose a VMOT $\pi$ for the perturbed cost $c_\delta (x_1, \dots, x_N) = c_1(x_1) + \delta x_{11} x_{12} + \sum_{t=2}^N c_N(x_N)$. 
  \footnote{The finite second moment assumption is used here to ensure the integrability of $x_{11} x_{12}$.}
  Note that the mixed derivative of $\delta x_1 x_2$ is $\delta$, implying that $c_1(x) + \delta x_1 x_2$ is strictly supermodular. Let $\pi$ be a VMOT for the cost $c_\delta$ for a fixed $\delta >0$. By part ii), $\pi^1 = \chi_{(\mu_{11}, \mu_{12})}$ is the monotone coupling of $\mu_{11}, \mu_{12}$. Let $\pi^{2, \dots, N} \in \cP(\R^{(N-1)d})$ denote the $(X_2,\dots,X_N)$-marginal of $\pi$. Now since $\pi$ is a VMOT, $\pi^{2, \dots, N}$ must  maximize $\mathbb{E}_\gamma[\sum_{i=2}^n c_i(X_i)]$ among all couplings $\gamma \in \bigcup_{(\mu_2, \dots, \mu_n) \in {\cal A}} \Pi(\mu_2, \dots, \mu_n)$, 
  where $(\mu_2, \dots, \mu_n) \in {\cal A}$ means that $\mu_t \in \Pi(\mu_{t,1}, \mu_{t,2})$ for $t = 2, \dots, N$ and satisfies the convex order $\chi_{(\mu_{11}, \mu_{12})} \preceq_c \mu_2 \preceq_c \dots \preceq_c \mu_N$.
  This implies that $\pi$ remains a VMOT for the cost $c_\delta (x_1, \dots, x_N)$ for all $\delta > 0$.
  Letting $\delta \searrow 0$, we conclude that $\pi$ remains a VMOT for the original cost $c(x_1, \dots, x_N)$.
\end{proof}

\section{Counterexamples and Related Examples}\label{sec:counterexamples_and_related_examples}

Theorem \ref{main} and the conjecture \ref{conjecture1} propose structural properties for VMOT under supermodular costs. 
However, these properties face limitations in certain scenarios. 
In this section, we present four examples to explore these boundaries and other related aspects of the VMOT framework. 

\subsection{Counterexamples to the Strict Supermodularity of the Cost}

In Theorem \ref{main}, if $c_1$ is not strictly supermodular, 
then the first marginal $\pi^1$ of a VMOT $\pi$ is not necessarily monotone. 
We now provide a counterexample to demonstrate this. 
Recall that $\conv(A)$ denotes the convex hull of a set $A$ in a vector space.

\begin{example}
  The strict supermodularity of $c_1$ is necessary for part ii) of Theorem \ref{main}. To construct a counterexample VMOT $\pi$ via duality, we take a convex function $\psi_1(y_1) = \frac13 |y_1|^3$ and its convex conjugate $\psi_2(y_2) = \psi_1^*(y_2) = \frac23 |y_2|^{\frac32}$. 
The inequality $\psi_1(y_1) + \psi_2(y_2) \ge y_1 y_2$ holds for all $y_1, y_2 \in \R$, and the set $\Gamma_{\{\psi_1, \psi_2\}}$ is defined as:
  \begin{align*}
    \Gamma_{\{ \psi_1,\psi_2\}} &:= \{(y_1,y_2) \in \R^2 \ | \ \psi_1(y_1) + \psi_2(y_2) = y_1y_2  \} \\
    & =  \{(y_1,y_2) \ | \ y_2 = \psi_1 ' (y_1) \} \nn \\
    & =  \{(y_1,y_2) \ | \ y_2 = |y_1|^2 \text { if } y_1 \ge 0, \ y_2 = -|y_1|^2 \text { if } y_1 \le 0 \}.  \nn
  \end{align*}
  
  Let $z = (-1, 1)$ and $w = (1, -1)$, and define the measure $\pi^1 = \frac12 \delta_z + \frac12 \delta_w \in \cP(\R^2)$. 
  Now, choose martingale kernels $\kappa_z, \kappa_w \in \cP(\R^2)$ such that:
  \[
  \int_{\R^2} x \, \kappa_z(dx) = z, \quad \int_{\R^2} x \, \kappa_w(dx) = w, \quad \kappa_z (\Gamma_{\{\psi_1, \psi_2\}}) = \kappa_w (\Gamma_{\{\psi_1, \psi_2\}}) = 1.
  \]
  This choice is possible because $\conv(\Gamma_{\{\psi_1, \psi_2\}}) = \R^2$.

  We then define a martingale measure $\pi$ such that the first marginal is $\pi^1$ and the kernel is $\{\kappa_z, \kappa_w\}$. 
  Now take the cost function $c(x, y) = y_1 y_2$, with $\phi_1 = \phi_2 = h_1 = h_2 = 0$, and note that $(\phi_i, \psi_i, h_i)_{i=1,2}$ and $\pi$ jointly satisfy the optimality conditions \eqref{ptwiseineq} and \eqref{ptwiseeq}. 
  This implies that $\pi$ is a VMOT in the class ${\rm VMT}(\mu_1, \mu_2, \nu_1, \nu_2)$, where $\mu_1, \mu_2, \nu_1, \nu_2$ are the one-dimensional marginals of $\pi$. 
  However, by construction, the first marginal $\pi^X = \frac12 \delta_z + \frac12 \delta_w$ is not monotone. 
\end{example}

\subsection{Counterexamples to the conjecture}
We demonstrate that the conjecture fails when $d > 2$. 
Consider the following examples:
  \begin{example}\label{conjectureisfalse}
  Conjecture \ref{conjecture1} is false if $d \ge 3$. 
  Specifically, there exist vectorial marginals $\vec \mu = ( \mu_1, \mu_2, \mu_3)$, $\vec \nu = ( \nu_1, \nu_2, \nu_3)$ satisfying $\mu_i \preceq_c \nu_i$, $i=1,2,3$,  
  such that for every VMOT $\pi$ to the problem \eqref{VMOT} with the cost $c = c(y) = y_1y_2+y_2y_3+y_3y_1$ and the marginals $\vec \mu, \vec \nu$, 
  its first time marginal $\pi^1$ fails to be the monotone coupling of $\vec \mu = (\mu_1, \mu_2, \mu_3)$.

  Let $n_0^+ = (0,0,1)$, $n_1^+ = (1,0,1)$, $n_2^+ = (0,1,1)$, $n_{12}^+ = (1,1,1)$, $n_0 = (0,0,0)$, $n_1 = (1,0,0)$, $n_2 = (0,1,0)$, $n_{12} = (1,1,0)$, $n_0^- = (0,0,-1)$, $n_1^- = (1,0,-1)$, $n_2^- = (0,1,-1)$, $n_{12}^- = (1,1,-1)$ 
  be the vertices of two vertically stacked cubes in $\R^3$, 
  and let $\cal Y \subset$ $\R^3$ be the set of these twelve vertices.  
Then, define $\bt_0 : \R^3 \to \R \cup \{ +\infty \}$ as:
  \begin{align*}
  &\bt_0 (n_0^+) = 0, \ \bt_0 (n_1^+) = 0, \  \bt_0 (n_2^+) = 0, \ \bt_0 (n_{12}^+) = 0, \\
  &\bt_0 (n_0) = 0, \ \bt_0 (n_1) = 1, \ \bt_0 (n_2) = 0, \ \bt_0 (n_{12}) = 1, \\
  &\bt_0 (n_0^-) = 0, \ \bt_0 (n_1^-) = 2, \ \bt_0 (n_2^-) = 1, \ \bt_0 (n_{12}^-) = 2,\\
  &\bt_0 = +\infty \ \text{ on } \ \R^3 \setminus \cal Y. 
  \end{align*}

  It is clear that $\bt_0$ is submodular, and moreover, $\bt_0^{**}$ is given by the supremum of three affine functions; $\bt_0^{**} = \max(L_1, L_2, L_3)$ in $\conv(\cal Y)$, where
  \begin{align*}
  &L_1 (y) = 0 \ \text{ for } \ y = (y_1,y_2,y_3) \in \R^3,\\
  &L_2 (y) = y_1 + y_2 - y_3 - 1,\\
  &L_3 (y) = 2y_1 - y_3 - 1.
  \end{align*}
  
  \tdplotsetmaincoords{70}{112} 
  \begin{figure}[h!]
    \centering
    \begin{minipage}[b]{0.49\textwidth} 
    \centering
    \begin{tikzpicture}[scale=1.8, tdplot_main_coords]
      
      \draw[thick,->] (0,0,0) -- (2,0,0) node[anchor=north east]{$y_1$};
      \draw[thick,->] (0,0,0) -- (0,1.5,0) node[anchor=north west]{$y_2$};
      \draw[thick,->] (0,0,0) -- (0,0,1.5) node[anchor=south]{$y_3$};
      
      \foreach \x/\y/\z/\val in {
        0/0/1/0, 1/0/1/0, 0/1/1/0, 1/1/1/0,
        0/0/0/0, 1/0/0/1, 0/1/0/0, 1/1/0/1,
        0/0/-1/0, 1/0/-1/2, 0/1/-1/1, 1/1/-1/2
      } {
        \filldraw (\x,\y,\z) circle (1pt) node[above right,xshift=2mm, yshift=-1mm] {\val};
      }
      
      \foreach \xstart/\ystart/\zstart/\xend/\yend/\zend in {
        0/0/0/1/0/0,
        0/1/0/1/1/0,
        0/0/1/1/0/1,
        0/1/1/1/1/1,
        0/0/-1/1/0/-1,
        0/1/-1/1/1/-1
      } {
        \draw (\xstart,\ystart,\zstart) -- (\xend,\yend,\zend);
      }
      
      \foreach \xstart/\ystart/\zstart/\xend/\yend/\zend in {
        0/0/0/0/1/0,
        1/0/0/1/1/0,
        0/0/1/0/1/1,
        1/0/1/1/1/1,
        0/0/-1/0/1/-1,
        1/0/-1/1/1/-1
      } {
        \draw (\xstart,\ystart,\zstart) -- (\xend,\yend,\zend);
      }
      
      \foreach \xstart/\ystart/\zstart/\xend/\yend/\zend in {
        0/0/0/0/0/1,
        1/0/0/1/0/1,
        0/1/0/0/1/1,
        1/1/0/1/1/1
      } {
        \draw (\xstart,\ystart,\zstart) -- (\xend,\yend,\zend);
      }
      
      \foreach \xstart/\ystart/\zstart/\xend/\yend/\zend in {
        0/0/0/0/0/-1,
        1/0/0/1/0/-1,
        0/1/0/0/1/-1,
        1/1/0/1/1/-1
      } {
        \draw (\xstart,\ystart,\zstart) -- (\xend,\yend,\zend);
      }
      
    \end{tikzpicture}
    \caption{$\beta_0$ values on the vertices.}
    \end{minipage}
    \begin{minipage}[b]{0.49\textwidth} 
    \centering
    \begin{tikzpicture}[scale=1.8, tdplot_main_coords]
      
      \draw[thick,->] (0,0,0) -- (2,0,0) node[anchor=north east]{$y_1$};
      \draw[thick,->] (0,0,0) -- (0,1.5,0) node[anchor=north west]{$y_2$};
      \draw[thick,->] (0,0,0) -- (0,0,1.5) node[anchor=south]{$y_3$};
      
      \foreach \x/\y/\z/\val in {
        0/0/1/1, 1/0/1/2, 0/1/1/0, 1/1/1/0,
        0/0/0/0, 1/0/0/2, 0/1/0/0, 1/1/0/1,
        0/0/-1/0, 1/0/-1/3, 0/1/-1/1, 1/1/-1/3
      } {
        \filldraw (\x,\y,\z) circle (1pt) node[above right,xshift=2mm, yshift=-1mm] {\val};
      }
      
      \foreach \xstart/\ystart/\zstart/\xend/\yend/\zend in {
        0/0/0/1/0/0,
        0/1/0/1/1/0,
        0/0/1/1/0/1,
        0/1/1/1/1/1,
        0/0/-1/1/0/-1,
        0/1/-1/1/1/-1
      } {
        \draw (\xstart,\ystart,\zstart) -- (\xend,\yend,\zend);
      }
      
      \foreach \xstart/\ystart/\zstart/\xend/\yend/\zend in {
        0/0/0/0/1/0,
        1/0/0/1/1/0,
        0/0/1/0/1/1,
        1/0/1/1/1/1,
        0/0/-1/0/1/-1,
        1/0/-1/1/1/-1
      } {
        \draw (\xstart,\ystart,\zstart) -- (\xend,\yend,\zend);
      }
      
      \foreach \xstart/\ystart/\zstart/\xend/\yend/\zend in {
        0/0/0/0/0/1,
        1/0/0/1/0/1,
        0/1/0/0/1/1,
        1/1/0/1/1/1
      } {
        \draw (\xstart,\ystart,\zstart) -- (\xend,\yend,\zend);
      }
      
      \foreach \xstart/\ystart/\zstart/\xend/\yend/\zend in {
        0/0/0/0/0/-1,
        1/0/0/1/0/-1,
        0/1/0/0/1/-1,
        1/1/0/1/1/-1
      } {
        \draw (\xstart,\ystart,\zstart) -- (\xend,\yend,\zend);
      }
      
    \end{tikzpicture}
    \caption{$\beta$ values on the vertices.}
    \end{minipage}
    \label{cube2}
  \end{figure}

  Define $\psi_1, \psi_2, \psi_3 : \R \to \R \cup \{+\infty\}$ by
  \begin{align*}
  &\psi_1(0) = 0, \ \psi_1(1) = 2, \ \psi_1 = +\infty \ \text{ else},\\
  &\psi_2(0) = 0, \ \psi_2(1) = 0, \ \psi_2 = +\infty \ \text{ else},\\
  &\psi_3(-1) = 0, \ \psi_3(0) = 0, \ \psi_3 (1) = 1, \ \psi_3 = +\infty \ \text{ else.}
  \end{align*}
  Set $\bt(y) = \sum_{i=1}^3 \psi_i(y_i) - c(y)$, where $c(y) = y_1y_2 + y_2y_3 + y_3y_1$. We have
  \begin{align*}
  &\bt (n_0^+) = 1, \ \bt (n_1^+) = 2, \  \bt (n_2^+) = 0, \ \bt (n_{12}^+) = 0, \\
  &\bt (n_0) = 0, \ \bt (n_1) = 2, \ \bt (n_2) = 0, \  \bt (n_{12}) = 1, \\
  &\bt (n_0^-) = 0, \ \bt (n_1^-) = 3, \ \bt (n_2^-) = 1, \ \bt (n_{12}^-) = 3,\\
  &\bt = +\infty \ \text{ on } \ \R^3 \setminus \cal Y.
  \end{align*}
  Notice $\bt \ge \bt_0$, and $\bt = \bt_0$ on ${\cal Z}:= \{n_0^-, n_2, n^+_{12} \}$. 
  Observed that $\bt_0 = \bt_0^{**}$ on ${\cal Z}$ (in fact also on $\cal Y$), 
  hence $\bt = \bt^{**}$ on ${\cal Z}$ as well. 
  We take $u = (\frac12, \frac34, \frac14)$, $u' = (\frac25, \frac45, \frac15)$ and their monotone rearrangement $\bar u = (\frac25, \frac34, \frac15)$, $\bar u' = (\frac12, \frac45, \frac14)$, 
  such that $\{u, u'\} \subset \conv({\cal Z})$, while none of $\bar u, \bar u'$ lies on the plane containing ${\cal Z}$.
  
  We now construct a vectorial martingale transport $\pi$. 
  For this, take $\pi^1 := \frac12(\delta_u + \delta_{u'})$ as the first time marginal of $\pi$. 
  Then we take the martingale kernel $\ka_1$ as the unique probability measure supported on $\cal Z$ with its barycenter $x$ in $\conv ({\cal Z})$. 
  Now define $\pi = \pi^1 \otimes \ka_1$ (note that we only need $\pi_1$ for $x=u, u'$), 
  and let $\vec \mu:= (\mu_1, \mu_2, \mu_3)$ be the 1D marginals of $\pi^1$, 
  and let $\vec \nu:= (\nu_1, \nu_2, \nu_3)$ be the 1D marginals of $\pi^2 = \frac12 (\ka_{u} + \ka_{u'})$. 
  We now claim that $\pi$ is a VMOT solving the problem \eqref{VMOT} with the cost $c$ and the marginals $\vec \mu, \vec \nu$. We will prove the optimality of $\pi$ by locating an associated dual optimizer $(\phi_i, \psi_i, h_i)_{i=1,2,3}$, 
  where $\psi_i$ has already been defined above. 
  Take $\phi_1 (x_1) = \frac12 x_1  $, $\phi_2(x_2) = \frac12 x_2$, $\phi_3(x_3) = - \frac12 x_3 - \frac12$, 
  so that $\sum_{i=1}^3 \phi_i =\frac{L_1+L_2}{2} = \frac12 x_1 + \frac12 x_2 - \frac12 x_3 - \frac12$. Now take $h(x) = (h_1(x), h_2(x), h_3(x))$ as 
  \be\label{h}
  h(x)=
  \begin{cases}
  \nabla L_1 = (0,0,0) \ \text{ if } \ L_1(x) > L_2(x),\\
  \nabla L_2 = (1,1,-1) \ \text{ if } \ L_1(x) < L_2(x),\\
  \nabla \frac{L_1+L_2}{2} = (\frac12, \frac12, -\frac12) \ \text{ if } \ L_1(x) = L_2(x).
  \end{cases}
  \ee
  In order to prove the optimality of $\pi$ and $(\phi, \psi, h)$ simultaneously, we need to confirm the optimality conditions \eqref{ptwiseineq} and \eqref{ptwiseeq}. To see \eqref{ptwiseineq}, observe
  \begin{align*}
  \sum_{i=1}^3 \phi_i(x_i) + h(x) \cdot(y-x) &= \bigg(\frac{L_1+L_2}{2}\bigg)(x) + h(x) \cdot(y-x) \\
  & \le \max(L_1, L_2) (y) \\
  & \le \bt_0^{**} (y) = \max(L_1, L_2, L_3) (y)\\
  & \le \bt^{**} (y) \\
  & \le \bt (y) = \sum_{i=1}^3 \psi_i (y_i) - c(y).
  \end{align*}
  Observe further that \eqref{ptwiseeq} follows by the fact that on $\cal Z$, 
  $L_3 \le L_1 = L_2 = \bt$, 
  such that the above inequalities become equalities for $x = u, u'$ and $y \in \cal Z$. 
  This simultaneously proves the optimality of $\pi$ and $(\phi_i, \psi_i, h_i)_{i}$ for the primal and dual problems respectively with the cost $c$ and the marginals $\vec \mu, \vec \nu$.
  
  Finally, take any $\ga \in {\rm VMT}(\vec \mu, \vec \nu)$, 
  such that its first time marginal $\ga^1 = \frac12(\delta_{\bar u} + \delta_{\bar u'})$ is monotone. 
  We claim that $\ga$ cannot be optimal. 
  If $\ga$ were optimal, it must satisfy the optimality condition \eqref{ptwiseeq} with the optimal dual $(\phi, \psi, h)$ constructed above. 
  However, we have $L_1 (\bar u) > L_2 (\bar u)$ and $L_1 (\bar u') < L_2 (\bar u')$, and this  implies the strict inequality
  \[
  \bigg(\frac{L_1+L_2}{2}\bigg)(x) + h(x) \cdot(y-x) < \max(L_1, L_2) (y) \ \text{ for } x = \bar u, \bar u' \text{ and } y \in \cal Y. 
  \]
  This shows that $\ga$ cannot satisfy \eqref{ptwiseeq}, and hence cannot be a VMOT for \eqref{VMOT}. 
  \end{example}

\subsection{Example for Tighter Bounds for Single-Period Costs in VMOT}

This example demonstrates how VMOT leverages its martingale structure to provide tighter bounds for costs that depend solely on a single period, outperforming classical OT in this setting.

\begin{example}\label{ex:uniform_vmot}
  Let $N = 2$ and $d = 2$, and consider $\mu_1 = \mu_2 = \nu_1$, which are uniform distributions on $[0.5, 1]$, while $\nu_2$ is uniform on $[0, 1.5]$. The cost function is $c(y_1, y_2) = e^{y_1 + y_2}$, depending solely on the second period. Notably, $c(y_1, y_2)$ is strictly supermodular.

  By Theorem \ref{main}, the first-period marginal coupling $\pi^1$ is monotone under any VMOT $\pi$. Since $\mu_1 = \mu_2$, the monotone coupling of these identical measures is realized by the identity map. Therefore, we conclude $X_1 = X_2$ almost surely under $\pi$. Moreover, as $\mu_1 = \nu_1$, the martingale constraint implies $X_1 = X_2 = Y_1$ almost surely under $\pi$.

  Let $(u_1,u_2,v_1,v_2,h_1,h_2)$ be a dual minimizer described in Theorem \ref{main2}. Then it holds
  \[
  u_1(x_1) + u_2(x_2) + v_1(y_1) + v_2(y_2) + h_1(x_1, x_2)(y_1 - x_1) + h_2(x_1, x_2)(y_2 - x_2) \geq c(y_1, y_2),
  \]
  where equality holds $\pi$-a.s.. Since $X_1 = X_2 = Y_1 = x$ a.s., the above dual constraint simplifies to
  \[
  u(x) + v(y) + h(x)(y - x) \geq c(x, y),
  \]
  where $Y_2 = y$, $u(x) = u_1(x) + u_2(x) + v_1(x), v(y) = v_2(y),$ and $h(x) = h_2(x, x)$. This simplified dual constraint implies that the VMOT problem reduces to a standard two-period MOT in $d = 1$.

  The cost function $c(x, y) = e^{x + y}$ satisfies the martingale Spence-Mirrlees condition $c_{xyy} > 0$, 
  ensuring that the optimal coupling for the second period is the left monotone coupling. \cite{bj, henry2016explicit}
  This results in a tighter upper bound for $\int c(y_1, y_2) \, d\pi$ compared to the monotone coupling between $\nu_1$ and $\nu_2$ in classical OT. 
  Specifically, the optimal VMOT cost is approximately 5.1124, while the classical OT cost is 5.2669.

  This example demonstrates how VMOT incorporates additional information, such as martingale constraints, to provide sharper bounds on the no-arbitrage prices (costs), offering insights into scenarios where multi-period dynamics influence optimal couplings.
\end{example}

\subsection{Example: Relaxing the Martingale Constraints: Marginal Martingale Transport}

We consider a relaxation of the VMOT problem, 
introducing the Marginal Martingale Optimal Transport (MMOT) framework. 
This framework simplifies the constraints by requiring marginal martingale conditions, 
explores a simpler coupling condition compared to the full vectorial martingale transport. 
The MMOT problem is defined as:
\[
  {\rm maximize } \ \ \E_\pi [c(X)] \ \text{ over } \ \pi \in {\rm MMT}(\vec \mu),
\]
where the feasible set ${\rm MMT}$ (Marginal Martingale Transport) consists of couplings $\pi$ satisfying:
\begin{align*}
  {\rm MMT}(\vec\mu) := \{ \pi \in \ &\cP(\R^{Nd}) \ | \  \pi = {\rm Law} (X), \, X = (X_1,\dots,X_N), \\
  &\E_\pi[X_{t+1,i}|X_{1,i},\dots X_{t,i}]=X_{t,i}, \, {\rm Law}(X_{t,i}) =\mu_{t,i}, \, \text{ for all } t \in [N], \, i \in [d] \}.
\end{align*}

\begin{example}
  To demonstrate the solution structure, we use the same setting as the previous example: $\mu_1 = \mu_2 = \nu_1$, which are uniform distributions on $[0.5, 1]$, and $\nu_2$ is uniform on $[0, 1.5]$. The cost function is $c(y_1, y_2) = e^{y_1 + y_2}$.

  To construct an optimal coupling $\pi \in \text{MMT}$, let $\gamma_1$ and $\gamma_2$ be any chosen martingale couplings of $\mu_1$ with $\nu_1$ and $\mu_2$ with $\nu_2$, respectively. ($X_1=Y_1$ under $\ga_1$.) 
  Disintegrate these couplings:
  \[
  \gamma_1 = \nu_1 \otimes \kappa_1^2, \quad \gamma_2 = \nu_2 \otimes \kappa_2^2,
  \]
  where $\kappa_1^2$ and $\kappa_2^2$ are conditional probability kernels.
  Define the coupling:
  \[
  \pi =  \pi_2 \otimes (\kappa_1^2 \otimes \kappa_2^2),
  \]
  where $\pi_2$ is the monotone coupling of $\nu_1$ and $\nu_2$.

  This construction ensures $\mathbb{E}_\pi[Y_i \mid X_i] = X_i$, satisfying the marginal martingale constraints. 
  Since the cost $c(y_1, y_2)$ depends only on the second period, the expected cost becomes:
  \[
  \mathbb{E}_\pi[c] = \mathbb{E}_{\pi_2}[c].
  \]
  Thus, $\pi_2$, being the monotone coupling, maximizes the cost and guarantees the optimality for the MMOT problem.

  Note that since $\chi_{\vec \mu} \not \preceq_c \chi_{\vec \nu}$, this example also illustrates that the MMOT problem and the VMOT problem can exhibit fundamentally different optimality properties.
\end{example}

\section{Numerical implementation}\label{Numerics}

Throughout this section, we assume that $d=2$ and the cost function $c$ is of the form $c(x_1, \dots, x_N) = \sum_{t=1}^N c_t(x_{t,1}, x_{t,2})$ where each $c_t$ is supermodular. We develop a variant of the VMOT problem that reduces its dimensionality taking advantage of our main result. We then employ a general version of the Sinkhorn algorithm, developed in \cite{hiew24} to compute the solutions for two examples in both general and reduced versions. As expected, the lower dimensional reduced version performs much better.

\subsection{A hybrid version of the VMOT Problem} \label{hybridVMOT}

Theorem \ref{main} motivates the following modified problem:
\begin{align}\label{monotone_VMOT}
\text{maximize} \quad \mathbb{E}_\pi [c(X)] \quad \text{over} \quad \pi \in {\rm VMT}(\chi,\vec\mu_2,\ldots,\vec\mu_N).
\end{align}
  
where 
\begin{align}\label{monotone_VMT}
{\rm VMT}(\chi,\vec\mu_2,\ldots,\vec\mu_N) := &\{ \pi \in \ \cP(\R^{2N}) \mid  \pi = {\rm Law} (X), \, X = (X_1,\dots,X_N), \\
&\;\;\E_\pi[X_{t} \mid X_1, \dots, X_{t-1}]=X_{t-1}, 
\, {\rm Law}(X_1) =\chi,
\, {\rm Law}(X_{t,i}) =\mu_{t,i}, \, \nn \\
&\;\text{ for all } t=2\ldots,N, \, i=1,2 \}, \nn
\end{align}
with some notation overload. This formulation assumes that we know the joint distribution of $X_1$. By Theorem \ref{main}, the VMOT problem \eqref{VMOT} is equivalent to the above problem in which $\chi \in \cP(\mathbb{R}^2)$ is the monotone coupling of $\vec\mu_1$, $\chi = \chi_{\vec \mu_1}$\footnote{Notice that in this case \eqref{monotone_VMT} refines the original definition in the sense that it defines a smaller search set that still contains the solution, suggesting an improvement in numerical efficiency.}. For fixed $\chi_{\vec\mu_1}$, we call $T$ the monotone map such that $\mu_{1,2} = T_\# \mu_{1,1}$, and $\tilde c(x_{1,1},x_2,\ldots,x_N)=c(x_{1,1}, T(x_{1,1}) ,\ldots,x_N)$. With this, we can rewrite \eqref{monotone_VMOT}  as 
\begin{align}\label{monotone_VMOT2}
\text{maximize} \quad \mathbb{E}_\pi [\tilde c(X_{1,1},X_2,\ldots,X_N)] \quad \text{over} \quad \pi \in {\rm VMT}(\chi_{\vec\mu_1},\vec\mu_2,\ldots,\vec\mu_N).
\end{align}

This problem is simpler than the original \eqref{VMOT}, due to the dimensional reduction in going from $X_1 \in \mathbb{R}^2$ to $X_{1,1}$ which is only 1 dimensional. We define the simplified space of functions
\begin{align*}
\tilde{\mathcal{H}} = \bigg\{ \tilde\varphi(x_{1,1},x_2,\ldots,x_N) &\coloneqq u_{1,1}(x_{1,1}) + h_{1,1}(x_{1,1})(x_{2,1} - x_{1,1}) + h_{1,2}(x_{1,1})(x_{2,2} - T(x_{1,1})) \nn \\ &\;\;\;\;\; +\sum_{t=2}^N\sum_{i=1}^2 u_{t,i}(x_{t,i})+ \sum_{t=2}^{N-1}\sum_{i=1}^2 h_{t,i}(x_{1,1}, x_2,\ldots,x_t)(x_{t+1,i}-x_{t,i})\nn \\
&\quad\quad \bigg| \ u_{t,i} \in L^1(\mu_{t,i}), \quad h_{t,i} \in L^\infty(\mathbb{R}^{td}), \quad t \in [N], \quad i=1,2 \bigg\}.
\end{align*}

The dual problem to \ref{monotone_VMOT2} is formulated as
\begin{equation}\label{monotone_dualproblem}
\inf_{(u,h)\in\tilde\Xi}\int u_{1,1} d\mu_{1,1}+\sum_{t=2}^N\sum_{i=1}^2\int u_{t,i}d\mu_{t,i}
\end{equation}
where $\tilde\Xi$ consists of functions $\tilde\varphi\in\tilde{\cal H}$ such that $\tilde\varphi(x_{1,1},x_2,\ldots,x_N)\ge \tilde c(x_{1,1}, x_2,\ldots,x_N)$. As with the primal problem, this dual is simpler than the original \eqref{dualproblem}, since we replace the two potential functions $u_{1,1},u_{1,2}$ with the single $u_{1,1}$, while the functions $h_{1,i}$ now take a scalar argument $x_{1,1}$ in place of $x_1$.

We can treat both the original VMT condition and the modified VMT condition as linear constraints over the set of couplings. For the original VMT condition, $\pi \in {\rm VMT}(\vec\mu)$ if and only if $\pi$ has the correct marginals and 
\begin{equation}\label{VMOTMartingaleConstraint}
\int_{\R^{Nd}} \sum_{t=1}^{N-1} \sum_{i=1}^2 h_{t,i}(x_1, \dots, x_t) (x_{t+1,i} - x_{t,i}) d \pi(x_1, \dots, x_t) = 0,
\end{equation}
for any $h_{i,t}$, while for the modified VMT condition, $\pi \in {\rm VMT}(\chi,\vec\mu_2,\ldots,\vec\mu_N)$ if and only if $\pi$ has the correct marginals and
\begin{align}
\int_{\R^{Nd}} &h_{1,1}(x_{1,1})(x_{2,1} - x_{1,1}) + h_{1,2}(x_{1,1})(x_{2,2} - T(x_{1,1})) \label{ModifyVMOTMartingaleConstraint}\\
& \qquad + \sum_{t=2}^{N-1} \sum_{i=1}^2  h_{t,i}(x_{1,1}, x_2, \dots, x_t) (x_{t+1,i} - x_{t,i}) d \pi(x_{1,1}, x_2, \dots, x_t) = 0 \nn,
\end{align}
for any $h_{i,t}$.

\subsection{Entropic regularization and Sinkhorn algorithm}

We next consider an entropic regularization of the VMOT problem, which enables efficient computation via a generalized Sinkhorn algorithm. Given $\vec\mu=(\mu_{t,i})_{t,i}$,  $\hat\mu = \otimes_{t,i}\mu_{t,i}$, and $\epsilon >0$, we solve
\be\label{eqn: regularized VMOT primal}
\text{maximize} \quad \int c\, d\pi - \varepsilon H_{\hat\mu}(\pi) 
\quad \text{over } \pi \in {\rm VMT}(\vec\mu),
\ee
where $H_{\hat\mu}(\pi)=\int \log(\frac{d\pi}{d\hat\mu})\,d\pi$ is the relative entropy. The corresponding dual is
\be\label{eqn: regularized VMOT dual}
\text{minimize} \quad 
\sum_{t,i}\int u_{t,i}\, d\mu_{t,i}
+ \varepsilon \int \exp\!\Big(\tfrac{c-\varphi}{\varepsilon}\Big)d\hat\mu,
\ee
where $\varphi(x) \coloneqq \sum_{t=1}^N \sum_{i=1}^2 u_{t,i}(x_{t,i}) + \sum_{t=1}^{N-1} \sum_{i=1}^2 h_{t,i}(x_1, \dots, x_t) (x_{t+1,i} - x_{t,i})$ as in \eqref{dualproblem}. Similarly, the reduced dimension problem \eqref{monotone_VMOT2} and its dual \eqref{monotone_dualproblem} may be regularized; this amounts to replacing $(x_{1,1},x_{1,2})$ by $(x_{1,1},T(x_{1,1}))$, removing $u_{1,2}$ and removing one dimension for each $h_{t,i}$ in \eqref{eqn: regularized VMOT primal} and \eqref{eqn: regularized VMOT dual}.  The regularized VMOT problems \eqref{VMOT} and \eqref{dualproblem} as well as their modified versions \eqref{monotone_VMOT2} and \eqref{monotone_dualproblem} fall into the class of optimal transport problems treated in \cite{hiew24}.  As $\varepsilon \rightarrow 0$, solutions to each of these problems converge to solutions of the unregularized versions; therefore, solving them for small $\varepsilon >0$ yields approximate solutions to the unregularized problems.  A general version of the celebrated Sinkhorn algorithm which applies to dual problems of this type was developed in \cite{hiew24}.  We will apply this algorithm to approximate solutions of \eqref{dualproblem} and \eqref{monotone_VMOT2}, illustrating the improved performance of the modified version \eqref{monotone_VMOT2}.


The generalized Sinkhorn algorithm alternates updates of the dual potentials $(u_{t,i})$ and the martingale multipliers $(h_{t,i})$.
Each update enforces the marginal and martingale constraints respectively while keeping other variables fixed.
The updates for $u_{t,i}$ admit closed-form exponential formulas, while those for $h_{t,i}$ are implicit and are solved numerically. Note that $h_{t,i}$ can be solved uniquely. 
Full derivations of the update rules are provided in Appendix~\ref{appendix: numerical details}.


Take the $d=2$, $N=2$ VMOT problem with each marginal discretized on $n$ points as an illustrative example; exploiting the monotone first–period coupling lowers the discrete unknowns from $n^4$ to $n^3$ in the reduced formulation. This dimensional reduction effect carries to the dual after entropic regularization, making each Sinkhorn step cheaper.
Empirically, for our three–period experiment in Table~\ref{tab:sinkhorn-3period}, the CPU time is reduced by $99\%$.

\subsection{Examples}

We illustrate the performance of the Sinkhorn‐based method and the dimensional reduction effect on two instances of the VMOT problem: a two‐period case and a three‐period case. 
All numerical calculations are performed in Python on a 13th Gen Intel(R) Core(TM) i7‐13620H 2.40\,GHz notebook.

\begin{figure}
    \centering
    \includegraphics[width=1.\linewidth]{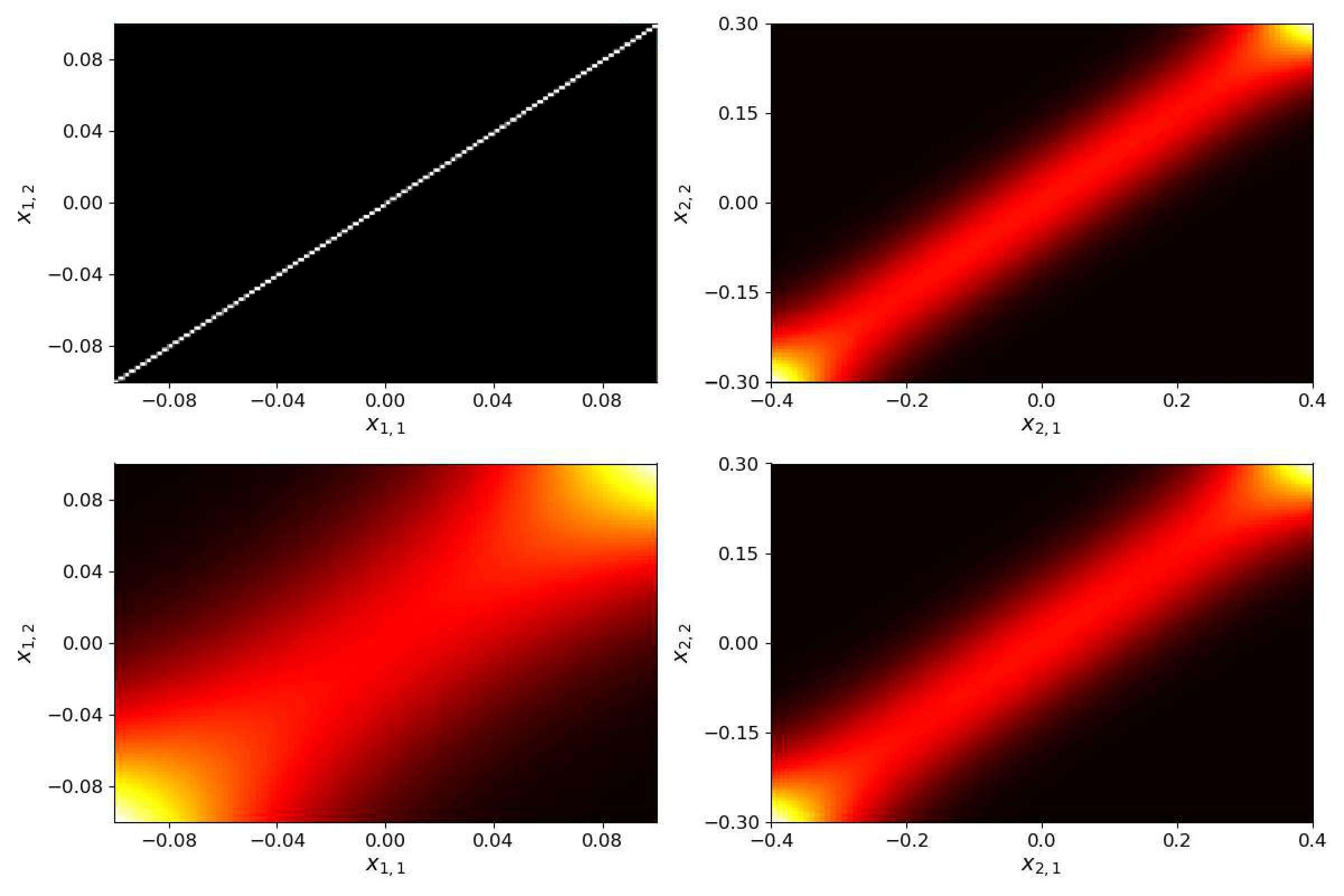}
    \caption{
    Heat maps of the induced couplings between different marginals obtained from the reduced (top row) and complete (bottom row) Sinkhorn algorithms. 
    Lighter colors indicate higher probability density. 
    }
    \label{plot2period}
\end{figure}

\paragraph{Two‐period case.}
Let $U(a,b)$ denote the uniform measure supported between $a$ and $b$.
Each marginal $\mu_{1,1} = U(-0.1,0.1), \mu_{1,2} = U(-0.1,0.1), \mu_{2,1} = U(-0.4,0.4), \mu_{2,2} = U(-0.3,0.3)$ is discretized on $100$ uniform grid points. 
The cost function is $c(x_1,x_2)=x_{1,1}x_{1,2}+x_{2,1}x_{2,2}$, for which each period is super-modular. The reduced Sinkhorn formulation achieves an optimal value of $0.02542$ in $2.52$ seconds, while the complete version yields $0.01906$ in $327.32$ seconds; note that the reduction in computation time is roughly two orders of magnitude, consistent with the heuristics described above for $n=100$. We can see the comparison in Table \ref{tab:sinkhorn-2period}. Note that it is not surprising that the reduced method achieves a higher optimal value, as it has an exact optimal coupling between $\mu_{1,1}$ and $\mu_{1,2}$ while the coupling deduced from the full Sinkhorn method is only approximately optimal (see the first column in Figure \ref{plot2period} for a comparison of these couplings).  Therefore, the total cost coming from the $x_{1,1}x_{1,2}$ term  in the cost function is necessarily higher in the reduced problem. Note as well that the optimal values reported in Table \ref{tab:sinkhorn-2period} do not include the entropy $H_{\hat\mu}(\pi)$ of the coupling.

\begin{table}[ht]
\centering
\caption{Two‐period VMOT problem ($d=2$, $N=2$). Each marginal discretized on $100$ grid points.}
\begin{tabular}{|c|c|c|}
\hline
\textbf{Method} & \textbf{Optimal value} & \textbf{Time (s)} \\ 
\hline
Full Sinkhorn & 0.01906 & 327.32 \\ 
\hline
Reduced Sinkhorn & 0.02542 & 2.52 \\ 
\hline
\end{tabular}
\label{tab:sinkhorn-2period}
\end{table}

We include the heat map plot in Figure \ref{plot2period} to demonstrate the induced calculated coupling between different marginals. Lighter colors represent higher probability and vice versa. The upper row is the plot for the reduced Sinkhorn method, while the bottom row is the complete Sinkhorn method. The first column is the plot for the marginals $\mu_{1,1}, \mu_{1,2}$, that is, the coupling for the first period, while the second column is the plot for the marginal $\mu_{2,1}, \mu_{2,2}$, i.e., the second period coupling. We see that all couplings are concentrated around the diagonal. For the first period, the complete Sinkhorn method demonstrates a slight deviation from the diagonal due to the error introduced by the entropic regularization, while the reduced Sinkhorn method concentrates right on the diagonal, which is the true coupling based on our previous result. This helps provide a better numerical result with less computational time.

\paragraph{Three‐period case.}
To test scalability, we add a third period with marginals $\mu_{3,1} = U(-0.6, 0.6),\mu_{3,2} = U(-0.7,0.7)$ discretized on $30$ grid points each, and $\mu_{1,1},\mu_{1,2},\mu_{2,1},\mu_{2,2}$ on $15$ grid points each. 
The cost function is extended as $c(x_1,x_2,x_3)=x_{1,1}x_{1,2}+x_{2,1}x_{2,2}+x_{3,1}x_{3,2}$.
The reduced method attains an optimal value of $0.006885$ in $57.28$ seconds, while the full formulation yields $0.006766$ in $15966.32$ seconds. 
In this example, we see that the reduction in computation time is very significant.

\begin{figure}
    \centering
    \includegraphics[width=1.\linewidth]{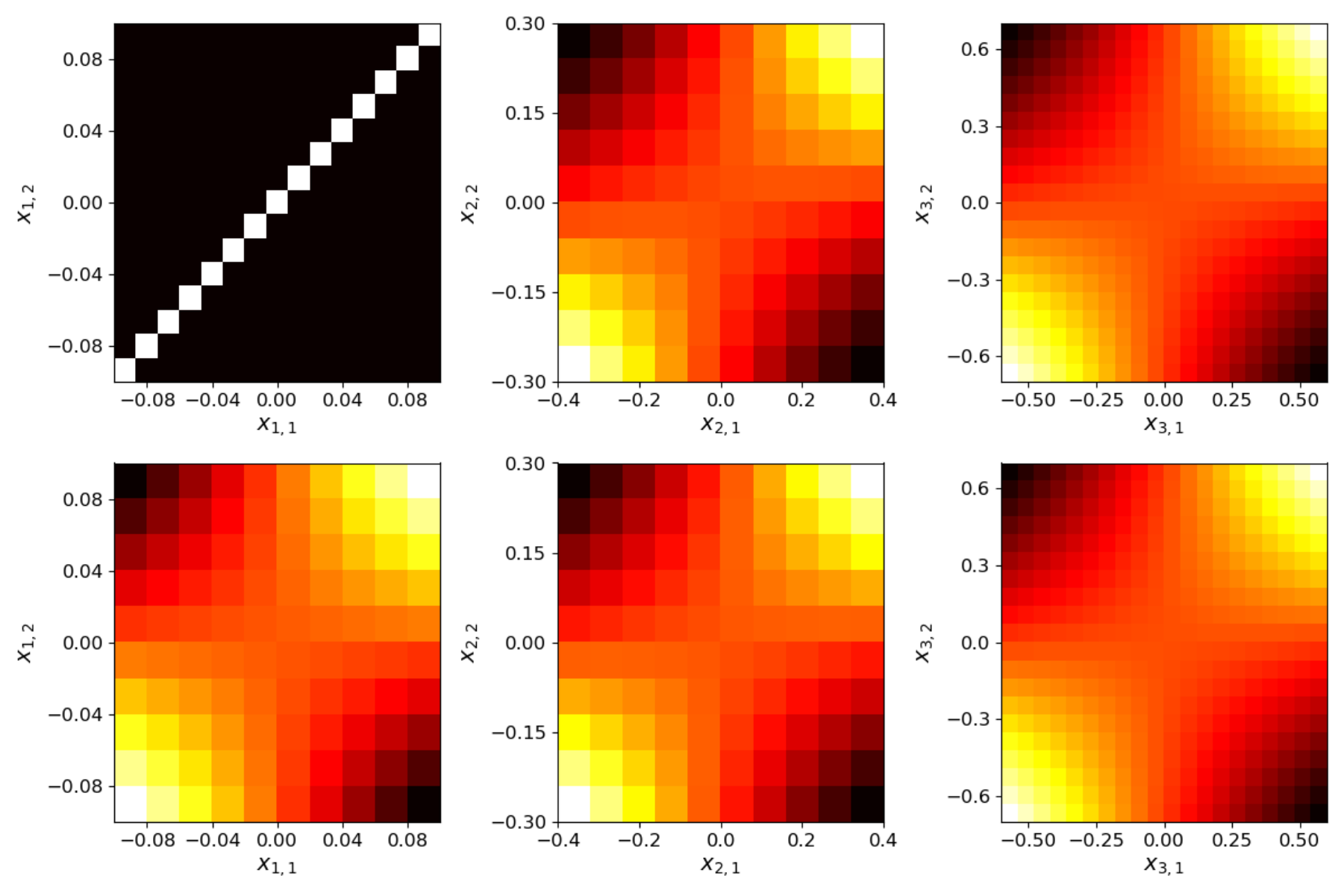}
    \caption{
    Heat maps of the induced couplings between different marginals obtained from the Sinkhorn algorithm for the three‐period case. 
    Lighter colors indicate higher probability density. 
    The figure compares the complete and reduced formulations discussed in the text.
    }
    \label{plot3period}
\end{figure}

\begin{table}[ht]
\centering
\caption{Three‐period VMOT problem ($d=2$, $N=3$). Marginals $\mu_{1,1},\mu_{1,2},\mu_{2,1},\mu_{2,2}$ discretized on $15$ grid points and $\eta_1,\eta_2$ on $30$ grid points.}
\begin{tabular}{|c|c|c|}
\hline
\textbf{Method} & \textbf{Optimal value} & \textbf{Time (s)} \\ 
\hline
Full Sinkhorn & 0.006766 & 15966.32 \\ 
Reduced Sinkhorn & 0.006885 & 57.28 \\ 
\hline
\end{tabular}
\label{tab:sinkhorn-3period}
\end{table}
We also include the heat map in Figure \ref{plot3period} for the three-period case. The top row is the plot for the reduced Sinkhorn method, while the bottom row is the complete Sinkhorn method. The first column is the plot for the marginal $\mu_{1,1}, \mu_{1,2}$. The second column is the plot for the marginal $\mu_{2,1}, \mu_{2,2}$. The third column is the plot for the marginal $\mu_{3,1}, \mu_{3,2}$. The plot also shows that the first period coupling for the reduced Sinkhorn method is significantly better than the complete Sinkhorn method.

\appendix

\section{Proof of Proposition \ref{prop1}}

\begin{proof}
  We begin by proving part i). If $\beta \equiv \infty$, then $\beta^* \equiv -\infty$, and there is nothing to prove. If $\beta \not\equiv \infty$ but $\beta^*$ is not proper (i.e., $\beta^* \equiv \infty$), again there is nothing to prove. Thus, we assume that $\beta$ and $\beta^*$ are proper.

  Suppose $\beta$ is submodular. For $R \geq 0$, define
  \[
  \beta_R(x) = \begin{cases}
  \beta(x) & \text{if } x \in [-R, R]^d, \\
  +\infty & \text{otherwise}.
  \end{cases}
  \]
  Note that $\beta_R$ is submodular, and for sufficiently large $R$, $\beta_R$ is proper. Since $\beta_R$ is compactly supported, $\beta_R^*$ will be Lipschitz unless $\beta_R^* \equiv +\infty$. However, the latter case is excluded because $\beta^* \geq \beta_R^*$. Therefore, $\beta_R^*$ is real-valued everywhere.

  Next, we show the supermodularity of $\beta_R^*$. By the definition of the Legendre transform, for any vectors $z, w \in \mathbb{R}^d$ and $\epsilon > 0$, there exists $x, y \in \mathbb{R}^d$ such that:
  \begin{align*}
    \beta_R^*(z) &\leq x \cdot z - \beta_R(x) + \epsilon, \\
    \beta_R^*(w) &\leq y \cdot w - \beta_R(y) + \epsilon,
  \end{align*}
  and
  \begin{align*}
    \beta_R^*(z \land w) &\geq (x \land y) \cdot (z \land w) - \beta_R(x \land y), \\
    \beta_R^*(z \lor w) &\geq (x \lor y) \cdot (z \lor w) - \beta_R(x \lor y).
  \end{align*}
  It is a direct calculation to show that:
  \[
  (x \land y) \cdot (z \land w) + (x \lor y) \cdot (z \lor w) \geq x \cdot z + y \cdot w,
  \]
  we can combine the inequalities to obtain:
  \begin{align*}
    \beta_R^*(z \land w) + \beta_R^*(z \lor w) &\geq (x \land y) \cdot (z \land w) + (x \lor y) \cdot (z \lor w) - \beta_R(x \land y) - \beta_R(x \lor y) \\
    &\geq x \cdot z + y \cdot w - \beta_R(x) - \beta_R(y) \\
    &= \beta_R^*(z) + \beta_R^*(w) - 2\epsilon.
  \end{align*}
  Taking the limit as $\epsilon \to 0$, we conclude that $\beta_R^*$ is supermodular.

  As $R \to \infty$, we have $\beta_R \searrow \beta$ on $\mathbb{R}^d$, and thus $\beta_R^* \nearrow \beta^*$. Therefore, the supermodularity of $\beta^*$ follows.

  Now we prove part ii). If $\beta(x, y)$ is supermodular on $\mathbb{R}^2$, then define $\tilde{\beta}(x, y) \coloneqq \beta(x, -y)$. By part i), $\tilde{\beta}$ is submodular, and thus $\tilde{\beta}^*$ is supermodular. Hence we have:
  \begin{align*}
    \tilde{\beta}^*(z, -w) &= \sup_{x, y} \left( x z + y (-w) - \tilde{\beta}(x, y) \right) \\
    &= \sup_{x, y} \left( x z - y w - \beta(x, -y) \right) \\
    &= \sup_{x, y} \left( x z + y w - \beta(x, y) \right) \\
    &= \beta^*(z, w),
  \end{align*}
   which shows that $\beta^*(z, w)$ is submodular. Part iii) follows from parts i) and ii).

\end{proof}

\section{Details of the numerical methods}\label{appendix: numerical details}

We illustrate the method when $N=2$ and $d=2$. For general $N$, the calculations are similar. First, we fix a discretization scheme for the marginals: $x_{1,1}^i, x_{2,1}^j, x_{1,2}^s, x_{2,2}^t$.

The dual problem of the entropic regularized VMOT problem with $N=2, d=2$ is 
\begin{align*}
\min_{u_{1,1}^i, u_{2,1}^j, u_{1,2}^s, u_{2,2}^t, h_1^{i,j}, h_2^{i,j}} &\sum_i u_{1,1}^i \mu_{1,1}^i + \sum_j u_{2,1}^j \mu_{2,1}^j + \sum_s u_{1,2}^s \mu_{1,2}^s + \sum_t u_{2,2}^t \mu_{2,2}^t\\
&\qquad+ \ep \sum_{i,j,s,t} \text{exp}(\frac{1}{\ep}(c^{i,j,s,t} - u_{1,1}^i - u_{2,1}^j - u_{1,2}^s-  u_{2,2}^t\\
&\qquad \qquad \qquad- h_1^{i,j}(x_{1,2}^s - x_{1,1}^i) - h_2^{i,j}(x_{2,2}^t - x_{2,1}^j))) \mu_{1,1}^i \mu_{2,1}^j \mu_{1,2}^s \mu_{2,2}^t.
\end{align*}

The generalized Sinkhorn method is an example of the block coordinate descent method \cite{hiew24}. The idea is to divide the set of variables to be optimized into different blocks and to optimize the objective function block by block while keeping all the other blocks fixed.

To implement the numerical algorithm, we first take the partial derivative of the dual functional with respect to each variable and set the derivative equal to 0. The resulting equations are then used to update the variables. We denote the dual functional by $\Phi(u_{1,1}^i, u_{2,1}^j, u_{1,2}^s, u_{2,2}^t, h_1^{i,j}, h_2^{i,j})$.

For the variables $u_{1,1}^i, u_{2,1}^j, u_{1,2}^s, u_{2,2}^t$, we take $u_{1,1}^i$ as an example, and the procedure for other variables is similar. If we take the partial derivative of $\Phi$ with respect to $u_{1,1}^i$ and we get
\begin{align*}
\frac{\partial \Phi}{\partial u_{1,1}^i} = \mu_{1,1}^i + \sum_{j,s,t} \text{exp}(\frac{1}{\ep}(&c^{i,j,s,t} - u_{1,1}^i - u_{2,1}^j - u_{1,2}^s-  u_{2,2}^t\\
&- h_1^{i,j}(x_{1,2}^s - x_{1,1}^i) - h_2^{i,j}(x_{2,2}^t - x_{2,1}^j))) \mu_{1,1}^i \mu_{2,1}^j \mu_{1,2}^s \mu_{2,2}^t = 0.
\end{align*}

Solving for $u_{1,1}^i$, we get the new $u_{1,1}^{i,k+1}$ after the iteration: 
$$u_{1,1}^{i,k+1} = -\ep \ln ( \sum_{j,s,t} \text{exp}(\frac{1}{\ep}(c^{i,j,s,t} - u_{2,1}^{j,k} - u_{1,2}^{s,k} - u_{2,2}^{t,k} -h_1^{i,j,k}(x_{1,2} - x_{1,1}) - h_2^{i,j,k}(x_{2,2} - x_{2,1}))) \mu_{2,1}^j \mu_{1,2}^s \mu_{2,2}^t).$$

For $h_1^{i,j}$, we differentiate $\Phi$ and get
\begin{align*}
\frac{\partial \Phi}{\partial h_1^{i,j}} = - \sum_{s,t} \text{exp}(\frac{1}{\ep}(&c^{i,j,s,t} - u_{1,1}^i - u_{2,1}^j - u_{1,2}^s-  u_{2,2}^t\\
&- h_1^{i,j}(x_{1,2}^s - x_{1,1}^i) - h_2^{i,j}(x_{2,2}^t - x_{2,1}^j))) (x_{1,2}^s - x_{1,1}^i) \mu_{1,1}^i \mu_{2,1}^j \mu_{1,2}^s \mu_{2,2}^t = 0.
\end{align*}

We cannot solve for $h_1^{i,j}$ analytically. However, we can still simplify the equation. Then the new $h_1^{i,j}$ after the iteration will be 
$$\sum_{s,t} \text{exp}(\frac{1}{\ep}(c^{i,j,s,t} - u_{1,1}^{i,k} - u_{2,1}^{j,k} - u_{1,2}^{s,k} - u_{2,2}^{t,k} - h_1^{i,j,k+1}(x_{1,2}^s - x_{1,1}^i) - h_2^{i,j,k}(x_{2,2}^t - x_{2,1}^j)))(x_{1,2}^s - x_{1,1}^i)  \mu_{1,2}^s \mu_{2,2}^t = 0.$$

A similar procedure applies to $h_2^{i,j}$.

\bibliographystyle{plain}
\bibliography{biblio}

@article{henry2016explicit,
	author = {Henry-Labord{\`e}re, Pierre and Touzi, Nizar},
	journal = {Finance and Stochastics},
	pages = {635--668},
	publisher = {Springer},
	title = {An explicit martingale version of the one-dimensional Brenier theorem},
	volume = {20},
	year = {2016}}

@article{AnsariLutkebohmertNeufeldSester22,
	author = {Ansari, Jonathan and L{\"u}tkebohmert, Eva and Neufeld, Ariel and Sester, Julian},
	journal = {arXiv preprint arXiv:2204.01071},
	title = {Improved robust price bounds for multi-asset derivatives under market-implied dependence information},
	year = {2022}}

@article{bhp,
	author = {Beiglb{\"o}ck, Mathias and Henry-Labordere, Pierre and Penkner, Friedrich},
	journal = {Finance and Stochastics},
	pages = {477--501},
	publisher = {Springer},
	title = {Model-independent bounds for option prices---a mass transport approach},
	volume = {17},
	year = {2013}}

@article{bj,
	author = {Mathias Beiglb{\"o}ck and Nicolas Juillet},
	doi = {10.1214/14-AOP966},
	journal = {The Annals of Probability},
	number = {1},
	pages = {42 -- 106},
	publisher = {Institute of Mathematical Statistics},
	title = {{On a problem of optimal transport under marginal martingale constraints}},
	url = {https://doi.org/10.1214/14-AOP966},
	volume = {44},
	year = {2016}}

@article{blo,
	author = {Mathias Beiglb{\"o}ck and Tongseok Lim and Jan Ob{\l}{\'o}j},
	doi = {10.3150/17-BEJ1015},
	journal = {Bernoulli},
	number = {3},
	pages = {1640 -- 1658},
	publisher = {Bernoulli Society for Mathematical Statistics and Probability},
	title = {{Dual attainment for the martingale transport problem}},
	url = {https://doi.org/10.3150/17-BEJ1015},
	volume = {25},
	year = {2019}}

@article{bnt,
	author = {Mathias Beiglb{\"o}ck and Marcel Nutz and Nizar Touzi},
	doi = {10.1214/16-AOP1131},
	journal = {The Annals of Probability},
	number = {5},
	pages = {3038 -- 3074},
	publisher = {Institute of Mathematical Statistics},
	title = {{Complete duality for martingale optimal transport on the line}},
	url = {https://doi.org/10.1214/16-AOP1131},
	volume = {45},
	year = {2017}}

@article{BreedenLitzenberger,
	author = {Breeden, Douglas T and Litzenberger, Robert H},
	journal = {Journal of business},
	pages = {621--651},
	publisher = {JSTOR},
	title = {Prices of state-contingent claims implicit in option prices},
	year = {1978}}

@article{Brenier87,
	author = {Brenier, Yann},
	journal = {CR Acad. Sci. Paris S{\'e}r. I Math.},
	pages = {805--808},
	title = {D{\'e}composition polaire et r{\'e}arrangement monotone des champs de vecteurs},
	volume = {305},
	year = {1987}}

@article{Caffarelli96,
	author = {Caffarelli, Luis},
	journal = {Partial differential equations and applications},
	pages = {29--35},
	title = {Allocation maps with general cost functions},
	volume = {177},
	year = {1996}}

@article{Carlier03,
	author = {Carlier, G.},
	fjournal = {Journal of Convex Analysis},
	issn = {0944-6532},
	journal = {J. Convex Anal.},
	mrclass = {49Q20 (49N15)},
	mrnumber = {2044434},
	mrreviewer = {Luigi De Pascale},
	number = {2},
	pages = {517--529},
	title = {On a class of multidimensional optimal transportation problems},
	volume = {10},
	year = {2003}}

@article{ConardDittmarGhysels13,
	author = {Conrad, Jennifer and Dittmar, Robert F. and Ghysels, Eric},
	doi = {https://doi.org/10.1111/j.1540-6261.2012.01795.x},
	journal = {The Journal of Finance},
	number = {1},
	pages = {85-124},
	title = {Ex Ante Skewness and Expected Stock Returns},
	url = {https://onlinelibrary.wiley.com/doi/abs/10.1111/j.1540-6261.2012.01795.x},
	volume = {68},
	year = {2013}}

@article{DolinskySoner14,
	author = {Dolinsky, Yan and Soner, H Mete},
	journal = {Probability Theory and Related Fields},
	number = {1-2},
	pages = {391--427},
	publisher = {Springer},
	title = {Martingale optimal transport and robust hedging in continuous time},
	volume = {160},
	year = {2014}}

@article{EcksteinGuoLimObloj21,
	author = {Eckstein, Stephan and Guo, Gaoyue and Lim, Tongseok and Ob{\l}{\'o}j, Jan},
	journal = {SIAM Journal on Financial Mathematics},
	number = {1},
	pages = {158--188},
	publisher = {SIAM},
	title = {Robust pricing and hedging of options on multiple assets and its numerics},
	volume = {12},
	year = {2021}}

@article{GaHeTo11,
	author = {A. Galichon and P. Henry-Labord{\`e}re and N. Touzi},
	doi = {10.1214/13-AAP925},
	journal = {The Annals of Applied Probability},
	number = {1},
	pages = {312 -- 336},
	publisher = {Institute of Mathematical Statistics},
	title = {{A stochastic control approach to no-arbitrage bounds given marginals, with an application to lookback options}},
	url = {https://doi.org/10.1214/13-AAP925},
	volume = {24},
	year = {2014}}

@article{GangboMcCann96,
	author = {Wilfrid Gangbo and Robert J. McCann},
	doi = {10.1007/BF02392620},
	journal = {Acta Mathematica},
	number = {2},
	pages = {113 -- 161},
	publisher = {Institut Mittag-Leffler},
	title = {{The geometry of optimal transportation}},
	url = {https://doi.org/10.1007/BF02392620},
	volume = {177},
	year = {1996}}

@article{GangboSwiech98,
	author = {Gangbo, Wilfrid and {\'S}wi{\k{e}}ch, Andrzej},
	journal = {Communications on Pure and Applied Mathematics: A Journal Issued by the Courant Institute of Mathematical Sciences},
	number = {1},
	pages = {23--45},
	title = {Optimal maps for the multidimensional Monge-Kantorovich problem},
	volume = {51},
	year = {1998}}

@article{Heinich2002,
	author = {Henri Heinich},
	doi = {https://doi.org/10.1016/S1631-073X(02)02341-5},
	issn = {1631-073X},
	journal = {Comptes Rendus Mathematique},
	number = {9},
	pages = {793-795},
	title = {Probl{\`e}me de Monge pour n probabilit{\'e}s},
	url = {https://www.sciencedirect.com/science/article/pii/S1631073X02023415},
	volume = {334},
	year = {2002}}

@book{HenryLabordere,
	author = {Henry-Labord{\`e}re, Pierre},
	publisher = {CRC Press},
	title = {Model-free hedging: A martingale optimal transport viewpoint},
	year = {2017}}

@article{hiew24,
	author = {Hiew, Joshua Zoen-Git and Nenna, Luca and Pass, Brendan},
	journal = {Numerische Mathematik},
	title = {An ordinary differential equation for entropic optimal transport and its linearly constrained variants},
	year = {2025},
volume={to appear},
url={https://doi.org/10.1007/s00211-025-01499-y},
}

@article{Hobson98,
	author = {Hobson, David G},
	journal = {Finance and Stochastics},
	pages = {329--347},
	publisher = {Springer},
	title = {Robust hedging of the lookback option},
	volume = {2},
	year = {1998}}

@article{Hobson11,
	author = {Hobson, David},
	journal = {Paris-Princeton lectures on mathematical finance 2010},
	pages = {267--318},
	publisher = {Springer},
	title = {The Skorokhod embedding problem and model-independent bounds for option prices},
	year = {2011}}

@article{JackwerthRubinstein96,
	author = {Jackwerth, Jens Carsten and Rubinstein, Mark},
	doi = {https://doi.org/10.1111/j.1540-6261.1996.tb05219.x},
	journal = {The Journal of Finance},
	number = {5},
	pages = {1611-1631},
	title = {Recovering Probability Distributions from Option Prices},
	url = {https://onlinelibrary.wiley.com/doi/abs/10.1111/j.1540-6261.1996.tb05219.x},
	volume = {51},
	year = {1996}}

@article{kellerer1972,
	author = {Kellerer, Hans G},
	journal = {Mathematische Annalen},
	pages = {99--122},
	publisher = {Springer},
	title = {Markov-komposition und eine anwendung auf martingale},
	volume = {198},
	year = {1972}}

@article{KimPass2014,
	author = {Kim, Young-Heon and Pass, Brendan},
	journal = {SIAM Journal on Mathematical Analysis},
	number = {2},
	pages = {1538--1550},
	publisher = {SIAM},
	title = {A general condition for Monge solutions in the multi-marginal optimal transport problem},
	volume = {46},
	year = {2014}}

@article{Levin99,
	author = {Levin, Vladimir},
	journal = {Set-Valued Analysis},
	number = {1},
	pages = {7--32},
	publisher = {Springer},
	title = {Abstract cyclical monotonicity and Monge solutions for the general Monge--Kantorovich problem},
	volume = {7},
	year = {1999}}

@article{Lim22,
	author = {Lim, Tongseok},
	journal = {Mathematical Programming},
	pages = {1--35},
	publisher = {Springer},
	title = {Geometry of vectorial martingale optimal transportations and duality},
	year = {2023}}

@article{Lim23,
	author = {Lim, Tongseok},
	journal = {arXiv preprint arXiv:2307.00807},
	title = {Replication of financial derivatives under extreme market models given marginals},
	year = {2023}}

@article{Lorentz53,
	author = {Lorentz, Georg Gunther},
	journal = {The American Mathematical Monthly},
	number = {3},
	pages = {176--179},
	title = {An inequality for rearrangements},
	volume = {60},
	year = {1953}}

@article{NutzStebeggTan20,
	author = {Nutz, Marcel and Stebegg, Florian and Tan, Xiaowei},
	journal = {Stochastic Processes and their Applications},
	number = {3},
	pages = {1568--1615},
	publisher = {Elsevier},
	title = {Multiperiod martingale transport},
	volume = {130},
	year = {2020}}

@article{PassVargasJimenez2021,
	author = {Pass, Brendan and Vargas-Jim{\'e}nez, Adolfo},
	journal = {Advances in Mathematics},
	title = {Monge solutions and uniqueness in multi-marginal optimal transport via graph theory},
	volume = {428},
	year = {2023}}

@article{Pass2011,
	author = {Pass, Brendan},
	journal = {SIAM Journal on Mathematical Analysis},
	number = {6},
	pages = {2758--2775},
	publisher = {SIAM},
	title = {Uniqueness and Monge solutions in the multimarginal optimal transportation problem},
	volume = {43},
	year = {2011}}

@article{talponenViitasaari2014,
	author = {Talponen, Jarno and Viitasaari, Lauri},
	journal = {Mathematics and Financial Economics},
	pages = {153--157},
	publisher = {Springer},
	title = {Note on multidimensional Breeden--Litzenberger representation for state price densities},
	volume = {8},
	year = {2014}}

@book{Villani09,
	author = {Villani, C{\'e}dric and others},
	publisher = {Springer},
	title = {Optimal transport: old and new},
	volume = {338},
	year = {2009}}

\end{document}